\documentclass[12pt,a4paper]{article}

\usepackage[utf8]{inputenc}
\usepackage[T1]{fontenc}
\usepackage{textcomp}
\usepackage{amsmath,amssymb,amsthm}
\usepackage{newtxtext}
\usepackage[scaled=0.92]{helvet}
\usepackage{newtxmath}
\usepackage{bm}
\usepackage{microtype}
\usepackage{xcolor}
\usepackage{setspace}
\usepackage{geometry}
\usepackage{changepage}

\usepackage{booktabs}
\usepackage{multirow}
\usepackage{array}
\usepackage{siunitx}
\usepackage{threeparttable}
\usepackage{tabularx}
\usepackage{longtable}
\usepackage{pdflscape}
\usepackage{makecell}
\usepackage{adjustbox}

\usepackage{graphicx}
\usepackage{subcaption}
\usepackage{float}
\usepackage{afterpage}
\usepackage[section]{placeins}
\usepackage{algorithm}
\usepackage{algorithmic}

\usepackage{enumitem}
\usepackage{titlesec}

\usepackage[
  backend=biber,
  style=authoryear,
  natbib=true,
  maxcitenames=2,
  mincitenames=1,
  sortcites,
  hyperref=true
]{biblatex}
\DeclareDelimFormat{nameyeardelim}{\addcomma\space}
\definecolor{GoogleBlue}{HTML}{1A73E8}
\usepackage[
  colorlinks=true,
  linkcolor=GoogleBlue,
  citecolor=GoogleBlue,
  urlcolor=GoogleBlue
]{hyperref}
\usepackage{cleveref}

\titleformat{\section}{\Large\sffamily\bfseries}{\thesection}{0.75em}{}
\titleformat{\subsection}{\large\sffamily\bfseries}{\thesubsection}{0.75em}{}
\titleformat{\subsubsection}{\normalsize\sffamily\bfseries}{\thesubsubsection}{0.75em}{}
\titleformat{\paragraph}[runin]{\normalsize\sffamily\bfseries}{\theparagraph}{0.75em}{}[.]
\titlespacing*{\section}{0pt}{2.0ex plus 0.5ex minus 0.2ex}{1.0ex}
\titlespacing*{\subsection}{0pt}{1.5ex plus 0.4ex minus 0.2ex}{0.7ex}
\titlespacing*{\subsubsection}{0pt}{1.2ex plus 0.3ex minus 0.1ex}{0.5ex}
\titlespacing*{\paragraph}{0pt}{1.0ex plus 0.2ex minus 0.1ex}{0.75em}

\usepackage{fancyhdr}
\fancypagestyle{plain}{%
  \fancyhf{}
  \fancyfoot[C]{\sffamily\small\thepage}

}

\usepackage[font=small,labelfont={bf,sf},labelsep=period,
  justification=justified,singlelinecheck=false]{caption}
\AtBeginEnvironment{subfigure}{\captionsetup{font=footnotesize}}
\renewenvironment{abstract}{%
  \par\begin{adjustwidth}{1.5em}{1.5em}
  \vspace{0.5ex}\hrule height 0.6pt\vspace{0.8ex}
  \small\sffamily
  \noindent\textbf{Abstract.}\enspace\ignorespaces
}{%
  \par\vspace{0.8ex}\hrule height 0.6pt
  \end{adjustwidth}\par
}

\renewcommand{\footnoterule}{%
  \kern-3pt
  \hrule width 0.35\textwidth height 0.4pt
  \kern 2.6pt
}

\setlist{nosep,topsep=5pt,itemsep=2pt,parsep=1pt,leftmargin=1.8em}
\makeatletter
\renewcommand{\fps@algorithm}{tbp}
\makeatother

\DeclareCiteCommand{\parencite}
  {\usebibmacro{prenote}\bibopenparen}
  {\bibhyperref{\usebibmacro{cite}}}
  {\multicitedelim}
  {\usebibmacro{postnote}\bibcloseparen}
\DeclareCiteCommand{\textcite}
  {\usebibmacro{prenote}}
  {\bibhyperref{\printnames{labelname}\setunit{\nameyeardelim}\printfield{labelyear}}}
  {\multicitedelim}
  {\usebibmacro{postnote}}
\let\citep\parencite
\let\citet\parencite

\newcommand{\RQ}[1]{\textbf{RQ#1}}
\newcommand{\Exp}[1]{\textsc{Exp-#1}}
\newcommand{\variant}[1]{\texttt{#1}}

\newtheorem{definition}{Definition}

\title{%
  \large\bfseries
  Decomposable Reward Modeling and Realistic Environment Design\\
  for Reinforcement Learning-Based Forex Trading%
}
\author{%
  \normalsize
  Nabeel Ahmad Saidd\\
  \small Dr. A.P.J. Abdul Kalam Technical University, Lucknow, India\\
  \href{mailto:nabeelahmadsaidd@gmail.com}{\texttt{nabeelahmadsaidd@gmail.com}}\\
}
\date{}

\begin{document}

\maketitle

\begin{abstract}
  Reinforcement learning (RL) for foreign exchange (Forex) trading requires realistic execution modeling, interpretable reward design, and action spaces that respect market constraints. Existing approaches often simplify these aspects, limiting policy transparency and practical relevance. This paper presents a modular RL framework integrating three components: (1) a friction-aware execution engine enforcing anti-lookahead semantics---observations at $\mathrm{close}_t$, execution at $\mathrm{open}_{t+1}$, and mark-to-market at $\mathrm{close}_{t+1}$---while modeling spread, commission, slippage, rollover financing, and margin-triggered liquidation; (2) a decomposable 11-component reward architecture with fixed, pre-specified weights and per-step logging, enabling attribution, ablation, and auditability; and (3) a 10-action discrete interface supporting scaling, reduction, closure, and reversal, with legal-action masking enforcing margin-aware feasibility during training and evaluation. Empirical evaluation on EURUSD training-split data reveals non-monotonic interactions among reward components, demonstrating that additional penalties do not consistently improve performance. The full reward configuration achieves a terminal training Sharpe ratio of 0.765 and cumulative return of 57.09\%. Comparing action spaces, the extended 10-action interface increases cumulative return but reduces Sharpe relative to a conservative 3-action variant, revealing a return--activity trade-off under a fixed training budget. Scaling-enabled strategies consistently reduce drawdowns, with the combined configuration delivering the strongest endpoint performance. Supplementary diagnostics on GBPUSD, USDJPY, and AUDUSD exhibit qualitatively similar trends. These findings demonstrate the framework's diagnostic value and underscore the importance of controlled, interpretable design in RL-based financial systems.

\end{abstract}
\clearpage
\section{Introduction}
\label{sec:introduction}

Reinforcement learning (RL) has emerged as a prominent paradigm for sequential decision-making in complex and uncertain environments, with successful applications ranging from games to continuous control in robotics \citep{mnih2015humanlevel,sutton2018reinforcement}. Financial trading is a demanding RL problem because decisions are sequential, outcomes are delayed, and market microstructure imposes strict feasibility constraints~\citep{moody2001learning,zhang2020deep,gould2013limit}. Foreign exchange (Forex) trading is particularly challenging. It operates continuously (24/5), is typically leveraged, and is sensitive to transaction costs and financing frictions, while return distributions are heavy-tailed, noisy, and regime dependent \citep{cont2001empirical}. These properties increase gradient variance, complicate credit assignment, and make policy evaluation sensitive to execution assumptions.

Despite substantial recent activity in applied RL-for-trading research—such as predictive signal extraction \citep{deng2016deep}, multi-asset portfolio allocation \citep{jiang2017deep}, and raw return optimization \citep{meng2019reinforcement,theate2021application}—three structural gaps continue to limit practical use. First, \emph{environment fidelity} is often insufficient. Trading performance depends on execution mechanics that are difficult to simulate but essential in leveraged markets \citep{zhang2020deep}. Many studies rely on simplified simulators that under-model or omit spread-embedded execution friction, dynamic financing costs, margin-based liquidation, and strict decision-execution timing \citep{carapuco2018reinforcement,rundo2019deep}. Such simplifications widen the sim-to-real gap and can produce policies that are profitable only because the simulator is permissive.

Second, \emph{reward design opacity} obscures the learning signal. Conventional approaches typically optimize a single scalar reward (for example, Profit and Loss (PnL), equity delta, or a risk-adjusted proxy such as the Sharpe ratio) \citep{moody2001learning,sharpe1994sharpe,theate2021application}. Although such objectives are easy to optimize, they make credit assignment less transparent and obscure whether observed gains arise from timing skill, reduced turnover, or leverage control. Without systematic decomposition and ablation, reward shaping remains difficult to analyze, tune, and reproduce \citep{ng1999policy}.

Third, \emph{action-space granularity} is often underspecified. Existing formulations frequently rely on coarse discrete spaces (for example, buy/sell/hold commands) \citep{theate2021application} or unconstrained continuous position sizing derived from deterministic policy gradients \citep{jiang2017deep,schulman2017proximal,haarnoja2018soft}. Coarse spaces do not represent practical operations such as scaling into positions, reducing exposure, or executing explicit reversals. Continuous formulations can express richer control, but they make hard feasibility constraints such as margin limits and position bounds less transparent unless they are enforced explicitly, for example through action masking in structured discrete domains \citep{huang2020closer}.

To address these challenges, this paper presents a system-level, open-source framework rather than a new RL algorithm. The proposed system integrates: (i) a Gymnasium-compatible Forex environment with strict anti-lookahead execution, friction and rollover modeling, and margin-aware risk controls; (ii) an 11-component decomposable reward architecture with fixed aggregation, state-dependent clipping, and transparent per-step component logging; and (iii) a legal-action-masked 10-action discrete space with explicit primitives for pyramiding, martingale-style scaling, reduction, closure, and reversal. All experiments are driven through configuration-based protocols so that experimental variants can be generated without ad hoc code changes.

\paragraph{Contributions.}
Our primary contributions are methodological and infrastructural rather than algorithmic:
\begin{itemize}[leftmargin=1.4em]
    \item \textbf{System-level execution environment:} We encode anti-lookahead timing, transaction frictions, rollover financing, margin constraints, and liquidation as simulator-enforced properties rather than post hoc corrections.
    \item \textbf{Reward and action interface design:} We introduce an 11-component reward system and a 10-action legality-masked interface, enabling component-level attribution and explicit position-management primitives that are not available in coarse buy/sell/hold formulations.
    \item \textbf{Controlled experimental protocol:} We demonstrate the framework through three experiment families (reward ablation, action-space comparison, and scaling analysis) on the EURUSD, with all variants generated via configuration rather than structural code changes.
    \item \textbf{Reproducibility infrastructure:} We provide a complete pipeline for deterministic artifact generation under fixed seeds, resolved configurations, isolated outputs, and leakage-free feature processing.
\end{itemize}

The empirical emphasis of this work is methodological robustness rather than state-of-the-art performance benchmarking. Reproducibility and experimental scope, including seeding and evaluation partitioning, are detailed in Section~\ref{sec:experiments}; interpretations focus on system behavior and learning dynamics rather than stochastic performance variation, consistent with cautionary guidance in trading-RL evaluation~\citep{zhang2020deep}.

\paragraph{Research questions.}
The empirical program is organized around three focused research questions:
\begin{enumerate}[leftmargin=1.8em, label=\textbf{RQ\arabic*.}]
    \item Which reward components contribute useful training signal, and does progressive penalty activation alter training-period risk-adjusted return relative to a profit-only baseline?
    \item Does increased action-space granularity change training efficiency and risk-adjusted learning relative to a coarse three-action discrete formulation?
    \item How do asymmetric position-scaling strategies (pyramiding vs. martingale-style scaling) influence training-period risk profiles, drawdown dynamics, and average scaling behavior?
\end{enumerate}

See Section~\ref{sec:experiments} for the canonical experimental setup, including evaluation partitioning and reproducibility controls.

The remainder of the paper is organized as follows. \Cref{sec:related_work} situates the framework within existing RL trading literature and identifies the research gap. \Cref{sec:methodology} details the environment, reward, and agent design. \Cref{sec:experiments} describes the evaluation protocol and reproducibility controls. \Cref{sec:results} presents the empirical findings, and \Cref{sec:discussion} discusses implications. \Cref{sec:future_work} then outlines limitations and forward directions before \Cref{sec:conclusion} concludes.
\section{Related Work}
\label{sec:related_work}

This section situates the manuscript within reinforcement learning (RL) for financial trading, with emphasis on where prior work succeeds and where methodological gaps remain for leveraged foreign exchange (Forex) environments. We organize the review by research theme to separate algorithmic advances from simulation assumptions and evaluation practice.

\subsection{Reinforcement Learning for Financial Trading: What Has Been Solved and What Has Not}
Early financial RL studies established the feasibility of direct objective optimization. \citep{moody2001learning} demonstrated recurrent RL for trading with Sharpe-oriented objectives, while \citep{bertoluzzo2012testing} compared alternative RL configurations in financial markets. Taken together, these studies show that the learning problem is tractable, but they do not by themselves resolve how execution costs, financing, or liquidation should be represented in a leveraged simulator.

The deep-learning phase expanded representational capacity in different directions. \citep{deng2016deep} emphasized learned financial signal representations, whereas \citep{jiang2017deep} formulated multi-asset portfolio allocation with deep policy learning. \citep{theate2021application} demonstrated an end-to-end trading workflow centered on return optimization, and \citep{liu2021finrl} emphasized reusable benchmark infrastructure. These contributions therefore address representation, allocation, workflow, and reproducibility respectively; they are complementary rather than interchangeable baselines. The remaining comparison problem is simulator fidelity: trading-RL reviews identify simplified execution assumptions as a recurring limitation \citep{zhang2020deep}, while market-microstructure research explains why order-book and transaction details matter for feasible execution \citep{gould2013limit}. The present work addresses that interface between learning and accounting rather than proposing another representation model.

To make this distinction explicit, \Cref{tab:related_work_comparison} contrasts representative studies on market scope, reward formulation, and action-space design.

\begin{table*}[!tbp]
\centering
\caption{Representative RL trading studies compared by market, reward formulation, and action-space design.}
\label{tab:related_work_comparison}
\begin{adjustbox}{width=\linewidth}
\begin{tabular}{@{}llcc@{}}
\toprule
\textbf{Study} & \textbf{Market} & \textbf{Reward} & \textbf{Actions} \\
\midrule
\citep{moody2001learning}
& General    
& Scalar (log return / Sharpe proxy) 
& Continuous (portfolio weights) \\

\citep{deng2016deep}
& Equities   
& Scalar (profit / return) 
& 3 discrete (buy/hold/sell) \\

\citep{jiang2017deep}
& Crypto     
& Scalar (portfolio return) 
& Continuous (portfolio allocation) \\

\citep{carapuco2018reinforcement}
& Forex      
& Scalar (profit-based) 
& 3 discrete (buy/sell/hold) \\

\citep{rundo2019deep}
& Forex (HF) 
& Scalar (PnL / return) 
& Discrete \\

\citep{theate2021application}
& Equities   
& Scalar (net return) 
& 3 discrete (buy/hold/sell) \\

\citep{liu2021finrl}
& Equities   
& Scalar (portfolio return / reward shaping) 
& Discrete (multi-asset actions) \\

\textbf{This work}               
& \textbf{Forex} 
& \textbf{Decomposable (11 components)} 
& \textbf{10 discrete actions} \\
\bottomrule
\end{tabular}
\end{adjustbox}
\end{table*}

\subsection{Environment Fidelity and Temporal Causality}
Simulator realism remains one of the most persistent bottlenecks for credible trading RL \citep{zhang2020deep}. The concern is not only the presence of a spread: carrying costs, maintenance-margin constraints, and liquidation rules change which policies are feasible in leveraged Forex. Thus, partial friction models can confound a learning comparison by changing the environment's admissible behavior rather than merely adding noise.

Temporal causality is a related but distinct issue. When observation and execution timing are not strictly separated, unintentional lookahead leakage can inflate backtest quality~\citep{zhang2020deep}. Execution fidelity determines what actions are economically possible; temporal fidelity determines what information is available when those actions are chosen. The present framework makes both contracts explicit.

\subsection{Reward Design and Risk-Aware Objectives}
Reward construction in trading RL is historically dominated by scalar objectives such as raw return, equity delta, or risk-adjusted proxies \citep{meng2019reinforcement,theate2021application}. These objectives are easy to optimize, but they do not reveal whether an agent learns profitable timing, excessive turnover, or leverage overuse. This is the motivation for treating reward terms as an auditable taxonomy rather than as an opaque sum.

Risk-aware formulations can improve objective alignment. Sharpe-ratio optimization \citep{moody2001learning,sharpe1994sharpe}, Sortino-style downside sensitivity \citep{sortino1994performance}, and broader tail-risk considerations motivated by heavy-tailed return evidence \citep{cont2001empirical} each target different risk dimensions. Yet these designs are often still collapsed into a single scalar at training time, obscuring component-level attribution and making it difficult to diagnose which term is responsible for a policy shift.

Reward shaping theory \citep{ng1999policy} provides formal guarantees only under potential-based transformations; finance-specific penalties such as drawdown, overtrading, and margin-utilization terms are generally not policy-invariant. Consequently, component-wise ablation is methodologically necessary, not optional, when claiming that auxiliary terms improve learning.

\subsection{Action-Space Modeling and Legality Constraints}
Action-space design spans minimal discrete interfaces and continuous allocation policies. Minimal interfaces improve optimization stability but under-model execution operations such as scaling, partial reduction, and explicit reversal. Continuous actor-critic approaches, including proximal policy optimization (PPO) \citep{schulman2017proximal} and soft actor-critic (SAC) \citep{haarnoja2018soft}, provide expressive control, but feasibility must still be enforced explicitly; otherwise, the richer action parameterization does not by itself solve margin or position constraints.

In constrained market simulators, legality-aware discrete control is a practical middle ground: actions remain interpretable while invalid operations can be excluded directly via masking. Prior evidence on invalid-action masking in policy-gradient settings \citep{huang2020closer} supports this direction, but adoption in trading environments remains inconsistent and often limited to interaction-time checks rather than both interaction and target computation.

\subsection{Algorithmic Stability and Replay Under Financial Non-Stationarity}
Value-based methods are useful baselines here because they combine discrete control with relatively transparent target construction. Deep Q-Networks (DQN) \citep{mnih2015humanlevel} provide the base value-learning formulation, while Double Deep Q-Networks (DDQN) \citep{hasselt2016deep} separate action selection from evaluation. Dueling and distributional variants \citep{wang2016dueling,bellemare2017distributional} extend representation and return modeling, but they address a different axis from the execution and reward-accounting questions studied here.

Financial environments introduce additional instability: regime shifts, sparse high-quality transitions, and temporal dependence in replay. Prioritized experience replay \citep{schaul2016prioritized} and Rainbow-style combinations \citep{hessel2018rainbow} address part of this challenge, while sequence models such as long short-term memory (LSTM) architectures \citep{fischer2018deep} capture temporal structure more explicitly. Nonetheless, these algorithmic advances do not eliminate simulator-induced bias when execution fidelity is weak.

\subsection{Forex-Specific Evidence and Evaluation Standards}
Forex-focused RL studies provide evidence that learned trading policies can produce useful signals in currency markets: \citep{carapuco2018reinforcement} study Forex directly, and \citep{rundo2019deep} study an FX high-frequency setting, while \citep{lucarelli2019deep} evaluate an ensemble strategy in automated stock trading. Their emphasis on predictive or profitability outcomes complements, but does not replace, explicit execution semantics and decomposable incentives. Because Forex is non-stationary, leveraged, and sensitive to regime shifts, results remain difficult to compare when market assumptions, execution models, and evaluation windows vary simultaneously.

More generally, trading RL evaluation still suffers from heterogeneous reporting standards \citep{zhang2020deep}. Modern practice requires joint reporting of return, volatility, drawdown, and turnover, plus calibration against rule-based baselines; these requirements are consistent with classical risk-return foundations in portfolio theory \citep{markowitz1952portfolio}. Multi-seed uncertainty quantification and stress testing remain under-reported across Forex RL literature, so single-seed results should be treated as descriptive rather than definitive.

\subsection{Novelty and Contributions}
The above review reveals a specific gap: prior studies usually improve \emph{one} axis (algorithm, reward objective, or benchmark tooling) while under-specifying the interaction among execution realism, legality-aware control, and decomposable incentives. This manuscript targets that joint gap.

Unlike \citep{deng2016deep} and \citep{theate2021application}, this work enforces a strict anti-lookahead execution contract with explicit fill and mark timing in a leveraged Forex setting. In contrast to \citep{carapuco2018reinforcement} and \citep{rundo2019deep}, this work models rollover financing, maintenance margin constraints, and forced liquidation as explicit simulator mechanics rather than implicit risk penalties. Unlike typical scalar-reward designs \citep{meng2019reinforcement,theate2021application}, this work introduces an 11-component reward architecture with deterministic per-step logging to support auditable ablation. In contrast to environments that treat invalid actions only at interaction time, this work applies legality masks consistently in both data collection and value-target computation \citep{huang2020closer}.

Therefore, the contribution is not a claim of a universally superior RL algorithm; it is a systems-level methodological contribution that couples realistic execution, transparent incentives, and reproducible experimentation into a single auditable framework.

\section{Methodology}\label{sec:methodology}

This section formalizes the end-to-end research pipeline, from causal data preparation to environment transition mechanics and mask-aware value learning. The methodology is a system-level framework built around standard value-based learners rather than a new optimization algorithm. To improve reproducibility, we provide explicit algorithmic specifications for (i) execution and portfolio updates, (ii) compositional reward assembly, and (iii) training-loop optimization.

\paragraph{Reading guide.}
The algorithms are presented in operational order rather than file-name order: the execution algorithm (Algorithm~\ref{alg:execution_portfolio_update}) constructs the transition, the reward algorithm (Algorithm~\ref{alg:reward_computation}) converts its trace into a learning signal, and the training algorithm (Algorithm~\ref{alg:mask_aware_training}) consumes that signal. Thus, the compiled numbering is Algorithm~1 for execution, Algorithm~2 for reward, and Algorithm~3 for training.

\subsection{Design Principles and System Overview}
\begin{figure}[tbp]
\centering
\includegraphics[width=\linewidth,height=0.8\textheight,keepaspectratio]{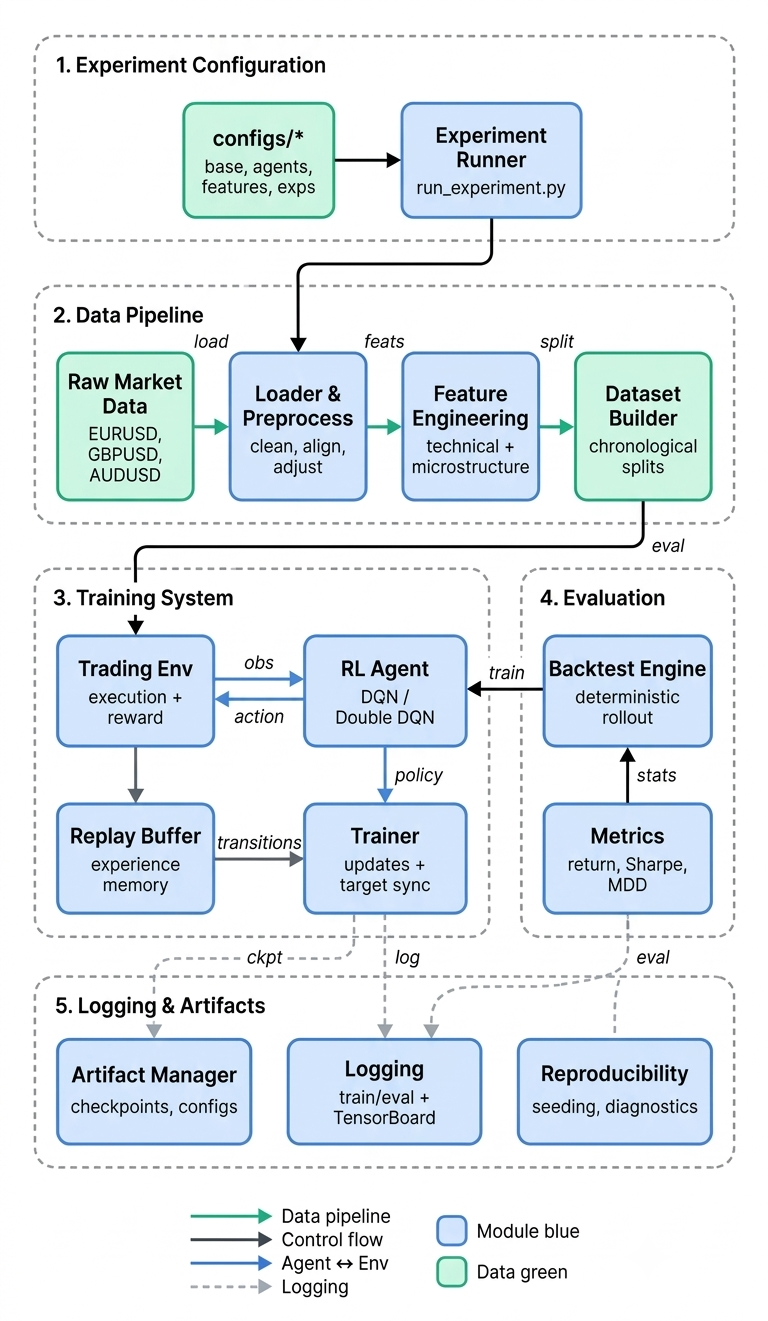}
\caption{System architecture of the RL trading framework, linking configuration, data processing, training, deterministic backtesting, and reproducibility logging in one end-to-end pipeline.}
\label{fig:system_architecture}
\end{figure}

The methodology is organized around three design principles: \emph{economic fidelity}, \emph{temporal causality}, and \emph{auditability}. Economic fidelity is enforced through explicit transaction-cost and margin accounting; temporal causality is enforced through strict chronological processing and anti-lookahead execution timing; and auditability is enforced through deterministic component ordering, configuration snapshots, and structured logging artifacts. The resulting end-to-end pipeline (data ingestion $\rightarrow$ preprocessing $\rightarrow$ environment simulation $\rightarrow$ value-function learning $\rightarrow$ standardized evaluation export) is summarized in \Cref{fig:system_architecture}.

\subsection{Data Pipeline, Temporal Partitioning, and Leakage Control}
The dataset consists of hourly open-high-low-close-volume (OHLCV) bars for EURUSD, GBPUSD, USDJPY, and AUDUSD (summary in \Cref{tab:dataset}). Data are normalized to Coordinated Universal Time (UTC), validated for schema integrity, deduplicated using a last-occurrence retention policy, and processed with deterministic missing-value handling.

\paragraph{Chronological processing policy.}
To preserve temporal causality, all experiments process data in strict chronological order with no shuffling. All evaluation metrics are computed exclusively on training trajectories and logged artifacts. No held-out or test split is used at any stage. This scope is deliberate: all reported results describe the training distribution, and no generalization or out-of-sample claims are made.

\paragraph{Scope and reproducibility.}
Feature transformations, warm-up handling, periodic evaluation, and final evaluation use only training data and stored training artifacts. The manuscript therefore reports deterministic, training-artifact-level results rather than held-out performance; reproducibility and seeding details are centralized in Section~\ref{sec:experiments}.

\paragraph{Train-only transformations and warm-up handling.}
Feature scaling parameters are estimated on the training data only and reused throughout. To initialize rolling indicators at the start of training, a warm-up lookback segment is prepended from the beginning of the training data, and warm-up rows are discarded from final training artifacts. These choices follow the scope and reproducibility policy above.

\subsection{Feature Engineering and Observation Construction}
Features are organized into deterministic namespaces so that all runs share a stable, auditable column ordering:
\begin{itemize}[leftmargin=1.4em]
    \item \textbf{Technical indicators}: simple and exponential moving averages (SMA/EMA; 10, 20, 50), relative strength index (RSI; 14), moving average convergence divergence (MACD; 12, 26, 9), Bollinger components (20, $2\sigma$), one-bar log return, and rolling volatility.
	\item \textbf{Microstructure proxies}: spread proxy $(H-L)/C$, short-horizon price-change proxy, realized volatility proxy, and UTC session label.
\end{itemize}
The canonical ordering improves reproducibility and prevents accidental feature-index drift across experiment variants.

\paragraph{Windowing and input dimensions.}
Let $L$ denote the observation lookback window ($L=24$ in release runs), $d_{\mathrm{feat}}$ the number of engineered market features, $d_{\mathrm{port}}=10$ the portfolio-state dimension, and $n_a$ the active action count ($n_a \in \{3,10\}$). The flattened MLP input dimension is
\begin{equation}
d_{\mathrm{flat}} = L\,d_{\mathrm{feat}} + d_{\mathrm{port}} + n_a.
\label{eq:flat_input_dim}
\end{equation}
where $d_{\mathrm{flat}}$ denotes the flattened input width, $L$ is the observation lookback window, $d_{\mathrm{feat}}$ is the number of engineered market features, $d_{\mathrm{port}}$ is the portfolio-state dimension, and $n_a$ is the active action count.

Equation~\ref{eq:flat_input_dim} is used consistently across training and evaluation to guarantee that architecture instantiation matches the declared observation schema.
The observation schema is summarized in \Cref{tab:obs_schema}.

\begin{table}[!tbp]
\centering
\caption{Observation schema and dimensions used by the Gymnasium environment.}
\label{tab:obs_schema}

\begin{adjustbox}{width=\linewidth}
\begin{tabularx}{\linewidth}{@{}l l X@{}}
\toprule
\textbf{Field} & \textbf{Shape} & \textbf{Role} \\
\midrule
\texttt{market} & $[L, d_{\mathrm{feat}}]$ & Rolling market-feature window ($L=24$) \\
\texttt{portfolio} & $[d_{\mathrm{port}}]$ & Normalized account state ($d_{\mathrm{port}}=10$) \\
\texttt{mask} & $[n_a]$ & Legal-action mask ($n_a\in\{3,10\}$) \\
\texttt{flat} & $[L d_{\mathrm{feat}} + d_{\mathrm{port}} + n_a]$ & Flattened vector for MLP agents \\
\bottomrule
\end{tabularx}
\end{adjustbox}
\end{table}

\subsection{Trading Environment}
The environment follows a Gym-style application programming interface (API), consistent with the OpenAI Gym design lineage \citep{brockman2016openai}, and exposes dictionary observations together with deterministic transition diagnostics. This explicit state contract enables direct reproducibility checks across training and evaluation workflows.

\subsubsection{Action Space and Legal Masking}

The extended action interface contains 10 operations:
\begin{itemize}
    \item \texttt{HOLD}
    \item \texttt{OPEN\_LONG}, \texttt{OPEN\_SHORT}
    \item \texttt{PYRAMID\_LONG}, \texttt{PYRAMID\_SHORT}
    \item \texttt{MARTINGALE\_LONG}, \texttt{MARTINGALE\_SHORT}
    \item \texttt{REDUCE}
    \item \texttt{CLOSE}
    \item \texttt{REVERSE}
\end{itemize}

For controlled comparisons, a simplified 3-action adapter is used:
\begin{itemize}
    \item \texttt{HOLD}
    \item \texttt{TARGET\_LONG}
    \item \texttt{TARGET\_SHORT}
\end{itemize}
These are mapped to valid extended operations conditioned on the current position state.

Action legality is recomputed at every step using direction, margin availability, and scaling-depth constraints. The legal mask is stored with transitions in replay and reused for both exploratory sampling and masked argmax target computation. Although prior empirical evidence is reported primarily for policy-gradient settings \citep{huang2020closer}, the same legality principle is applied here to value-based updates. \Cref{tab:action_definitions} summarizes extended-action semantics.
\begin{table*}[!tbp]
\centering
\caption{Extended action semantics and legality conditions (executed at $open_{t+1}$).}
\label{tab:action_definitions}

\begin{tabularx}{\linewidth}{@{}r l >{\raggedright\arraybackslash}X >{\raggedright\arraybackslash}X@{}}
\toprule
\textbf{ID} & \textbf{Action} & \textbf{Primary legality precondition} & \textbf{Execution effect} \\
\midrule
0 & HOLD & Always legal & Maintain current position; no new order submitted \\
1 & OPEN\_LONG & Flat position and sufficient free margin & Open long exposure at base lot size \\
2 & OPEN\_SHORT & Flat position and sufficient free margin & Open short exposure at base lot size \\
3 & PYRAMID\_LONG & Existing long position, scaling depth below cap, margin available & Add to long position using configured pyramid increment \\
4 & PYRAMID\_SHORT & Existing short position, scaling depth below cap, margin available & Add to short position using configured pyramid increment \\
5 & MARTINGALE\_LONG & Existing long position, martingale depth below cap, margin available & Add long exposure after adverse movement using martingale scaling rule \\
6 & MARTINGALE\_SHORT & Existing short position, martingale depth below cap, margin available & Add short exposure after adverse movement using martingale scaling rule \\
7 & REDUCE & Non-flat position & Partially reduce current exposure (configured reduction fraction) \\
8 & CLOSE & Non-flat position & Fully close current position \\
9 & REVERSE & Non-flat position and sufficient margin for opposite side & Close current direction, then open opposite direction within one action \\
\bottomrule
\end{tabularx}
\end{table*}

\subsubsection{Execution Timing and Market Frictions}
To prevent lookahead leakage, step timing is fixed as:
\begin{equation}
\begin{aligned}
\mathrm{observe}\,\mathrm{close}_t \;\rightarrow\; \mathrm{decide}\,a_t \;\rightarrow\; \mathrm{fill}\,\mathrm{open}_{t+1} \\
\rightarrow\; \mathrm{mark}\,\mathrm{close}_{t+1}.
\end{aligned}
\label{eq:execution_timing_contract}
\end{equation}
where $\mathrm{close}_t$ denotes the close price observed at step $t$, $a_t$ is the action selected after observing that close, $\mathrm{open}_{t+1}$ is the next bar's open used for execution, and $\mathrm{close}_{t+1}$ is the next bar's close used for mark-to-market.

\begin{figure}[tbp]
\centering
\includegraphics[width=\linewidth,height=0.8\textheight,keepaspectratio]{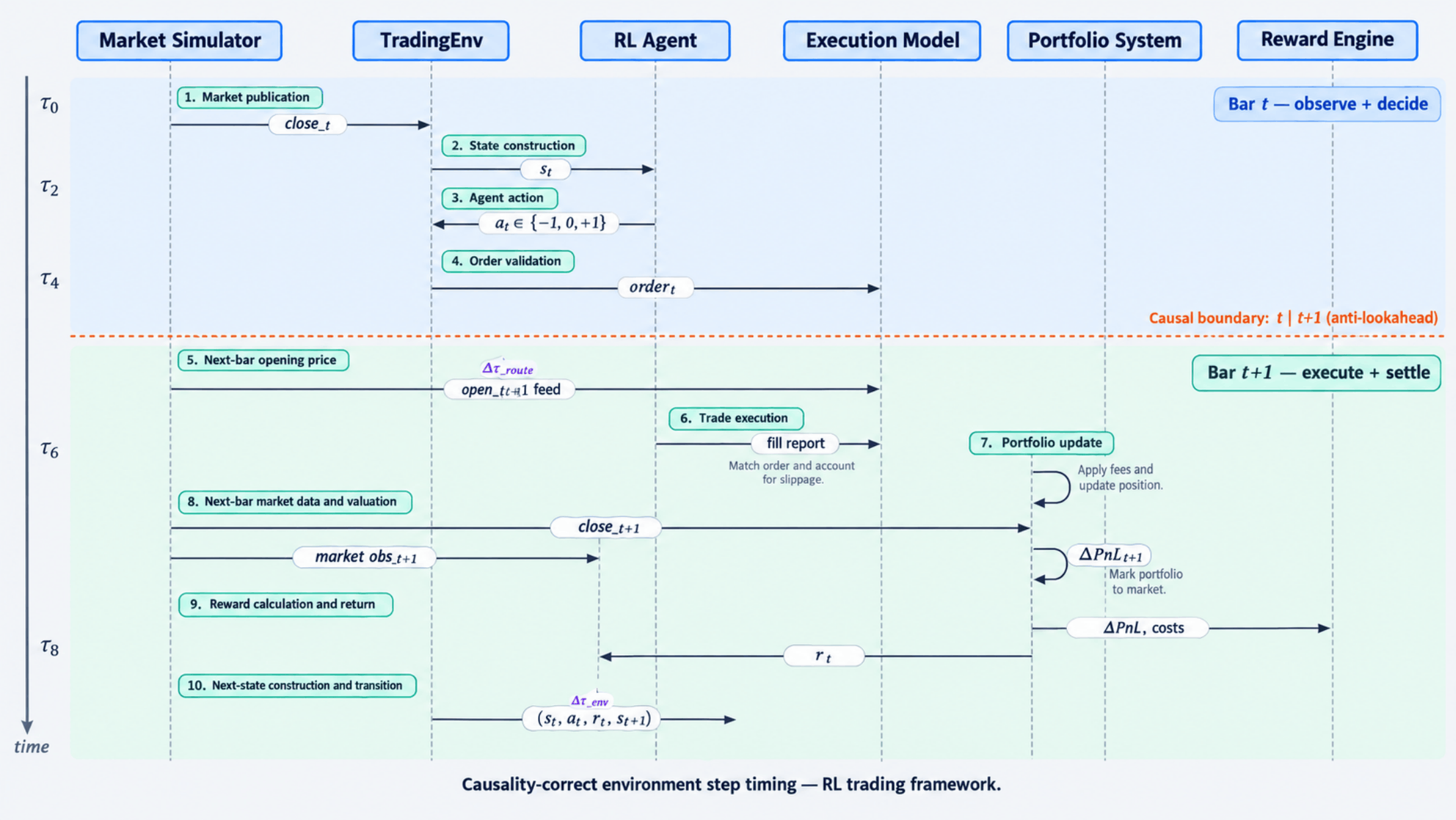}
\caption{Causality-correct environment step timing: the environment emits $\mathbf{s}_t$, receives $a_t$, execution is driven by market $\mathrm{open}_{t+1}$, portfolio state is marked at $\mathrm{close}_{t+1}$, reward engine returns only $r_t$ to the environment, and the environment emits the full transition $(\mathbf{s}_t,a_t,r_t,\mathbf{s}_{t+1})$ without any Reward$\rightarrow$Agent bypass.}
\label{fig:timing_diagram}
\end{figure}

The fill model applies spread, commission, and slippage in deterministic order; rollover financing is charged at 22:00~UTC with triple rollover on Wednesdays. Equation~\ref{eq:execution_timing_contract} is the causal timing contract summarized in \Cref{fig:timing_diagram}.

\subsubsection{Portfolio and Risk Accounting}
The account model tracks cash, equity, realized and unrealized PnL, used/free margin, directional exposure, scaling depth, and drawdown. Core constraints include maximum leverage ($30\times$), maintenance margin ($50\%$ of initial margin requirement), bounded scaling depth, and forced liquidation at a fixed equity threshold. These controls are necessary to prevent value-function optimization from exploiting infeasible leverage trajectories.

The execution algorithm fixes the operational order of legality checks, fill construction, cost deduction, rollover financing, mark-to-market, margin updates, and forced liquidation. Making this order explicit is necessary for economic fidelity: the transition must reflect a single deterministic accounting sequence.

\begin{algorithm}[!tbp]
\caption{Execution and portfolio update with frictions.}
\label{alg:execution_portfolio_update}
\begin{algorithmic}[1]
\STATE \textbf{Input:} portfolio state $P_t$, action proposal $a_t$, bars
\STATE \hspace*{1em} $(\mathrm{close}_t,\mathrm{open}_{t+1},\mathrm{close}_{t+1})$, configuration $\Omega$, legal mask $m_t$
\STATE \textbf{Output:} next portfolio state $P_{t+1}$, realized/unrealized PnL,
\STATE \hspace*{1em} cost trace, violation flag, and liquidation flag
\STATE Recompute $m_t$ from current direction, free margin, and scaling-depth limits
\IF{$m_t[a_t]=0$}
    \STATE Set executed action $\tilde{a}_t \leftarrow \texttt{HOLD}$ and raise constraint-violation flag
\ELSE
    \STATE Set $\tilde{a}_t \leftarrow a_t$
\ENDIF
\STATE Perform pre-trade feasibility checks for $\tilde{a}_t$ (margin and leverage)
\IF{pre-trade feasibility fails}
    \STATE Set $\tilde{a}_t \leftarrow \texttt{HOLD}$ and raise constraint-violation flag
\ENDIF
\STATE Set base execution anchor $p_{\mathrm{base}} \leftarrow \mathrm{open}_{t+1}$
\STATE Apply spread adjustment to obtain side-aware quoted price
\STATE Apply deterministic slippage to obtain final fill price $p_{\mathrm{fill}}$
\STATE Execute action semantics for $\tilde{a}_t$ (open, scale, reduce, close, reverse, or hold)
\STATE Deduct commissions from cash according to traded volume
\IF{timestamp equals rollover time (22:00 UTC)}
    \STATE Apply rollover financing (triple rollover on Wednesday)
\ENDIF
\STATE Mark open positions to $\mathrm{close}_{t+1}$ and update unrealized and realized PnL
\STATE Update cash, equity, used/free margin, utilization ratio, and drawdown statistics
\IF{equity below liquidation threshold \textbf{or} maintenance-margin rule violated}
    \STATE Force-liquidate all open positions using configured liquidation convention
    \STATE Set liquidation event flag and apply liquidation accounting effects
\ENDIF
\STATE Construct and return $P_{t+1}$ with per-step diagnostic traces
\end{algorithmic}
\end{algorithm}

Algorithm~\ref{alg:execution_portfolio_update} returns the next portfolio state together with realized and unrealized PnL, costs, and diagnostic flags. The transition trace is not yet a learning signal; it records realized PnL, costs, violation flags, and liquidation diagnostics that the reward algorithm converts into a scalar reward. This deliberate separation keeps execution accounting deterministic while leaving reward composition ready for ablation.

\subsection{Compositional Reward Modeling}

We model the reward as a weighted composition of components. Let $c_i(s_t, a_t, s_{t+1})$ denote component $i$ with weight $w_i$. The pre-normalized reward is:
\begin{equation}
R_t^{\text{raw}} = \sum_{i=1}^{11} w_i\, c_i(s_t, a_t, s_{t+1}).
\label{eq:raw_reward_sum}
\end{equation}
where $R_t^{\text{raw}}$ is the unbounded step reward, $w_i$ is the fixed weight assigned to component $i$, $c_i(s_t,a_t,s_{t+1})$ is the value of component $i$ computed from the transition, and $i$ indexes the 11 reward terms in Table~\ref{tab:reward_components}.

The final reward applies clipping:
\begin{equation}
R_t = \mathrm{clip}(R_t^{\text{raw}}, r_{\min}, r_{\max}).
\label{eq:reward_clip}
\end{equation}
where $R_t$ is the clipped reward used for temporal-difference updates, $r_{\min}$ and $r_{\max}$ are the lower and upper clipping bounds, and $\mathrm{clip}(\cdot)$ truncates its argument to the interval $[r_{\min}, r_{\max}]$. In the reward algorithm and training loop, this same clipped scalar is denoted by $r_t$; $R_t^{\mathrm{raw}}$ and $R_t$ are the prose/equation notation for the raw and clipped quantities, respectively.

Equation~\ref{eq:raw_reward_sum} defines deterministic component aggregation, and Equation~\ref{eq:reward_clip} bounds temporal-difference target magnitude.

\paragraph{Clipping vs. running normalization.}
Clipping bounds temporal-difference targets and stabilizes early training by limiting extreme updates. It preserves sign information and relative ordering within the unclipped region, but compresses large-magnitude rewards. In contrast, running normalization mitigates scale drift but introduces non-stationarity into the learning objective. We therefore adopt clipping as the default and use adaptive normalization only in sensitivity analyses.

\paragraph{Component structure and weighting.}
The 11 components span four functional groups: (i) return, represented by post-cost profit; (ii) risk control, represented by volatility and drawdown penalties; (iii) trading frictions, represented by transaction burden and overtrading penalties; and (iv) constraints, represented by margin usage, liquidation, and rule-violation terms. The table below summarizes the exact weights and behavioral roles.

\begin{table*}[!tbp]
\centering
\caption{Reward component definitions, baseline weights, and intended behavioral effects.}
\label{tab:reward_components}

\begin{adjustbox}{width=\linewidth}
\begin{tabular}{l r p{4.2cm} p{6.2cm}}
\toprule
\textbf{Component} & \textbf{Weight} & \textbf{Primary purpose} & \textbf{Typical trigger / penalty form} \\
\midrule
Profit (post-cost) & 1.00 & Core economic learning signal & Step-wise post-cost equity return \\
Holding incentive & 0.03 & Reward sustained profitable exposure & Small bonus when an open position remains profitable under contained drawdown \\
Volatility penalty & 0.01 & Discourage unstable equity trajectories & Negative term proportional to short-horizon equity-return volatility \\
Drawdown penalty & 0.05 & Penalize peak-to-trough equity loss & Incremental drawdown penalty with a stronger regime beyond the severe-drawdown threshold \\
Transaction burden & 0.10 & Reduce cost-driven churn & Cost-normalized penalty for spread, slippage, commission, and rollover \\
Overtrading penalty & 0.02 & Suppress unnecessary repositioning & Bounded penalty tied to elevated local trade frequency \\
Pyramiding penalty & 0.05 & Constrain aggressive winner-adding & Depth-aware penalty for additional pyramid levels \\
Martingale penalty & 0.12 & Discourage averaging down under leverage & Depth-aware penalty with a larger coefficient than pyramiding \\
Margin-utilization penalty & 0.05 & Avoid excessive leverage before liquidation & Nonlinear penalty when utilization exceeds a conservative threshold \\
Liquidation penalty & 2.00 & Impose strong aversion to forced liquidation & Large penalty on a forced-liquidation event \\
Constraint-violation penalty & 0.10 & Deter invalid action proposals & Penalty when the policy attempts an invalid action \\
\bottomrule
\end{tabular}
\end{adjustbox}
\end{table*}

The economic and risk components have distinct roles. Profit is the primary economic signal and rewards post-cost equity growth. Holding incentive rewards sustained profitable exposure when the position remains favorable and drawdown is contained. Volatility and drawdown penalties discourage unstable equity trajectories and large peak-to-trough losses.

The friction and constraint components regulate how that exposure is obtained. Transaction burden and overtrading penalties reduce churn by discouraging spread, slippage, commission, rollover, and unnecessary re-entry. Pyramiding and martingale penalties distinguish controlled scale-in after favorable movement from averaging down into adverse movement; the martingale term is weighted more heavily because loss-averaging under leverage is riskier than adding exposure after gains. Margin-utilization, liquidation, and constraint-violation penalties encode broker-style feasibility limits and penalize states or actions that threaten forced exit. The weights are a calibrated baseline for these roles, not a universal optimum.

The baseline weights were fixed before the main experiments so that auxiliary terms regularize the policy without displacing profit as the dominant signal. This choice also makes the reward easier to interpret: the ablation family in Section~\ref{sec:results} serves as a sensitivity check showing how changes in component activation alter the return--drawdown trade-off. The reported weights should therefore be read as a calibrated baseline rather than as a universally optimal setting.

\begin{figure}[tbp]
\centering
\includegraphics[width=\linewidth,height=0.8\textheight,keepaspectratio]{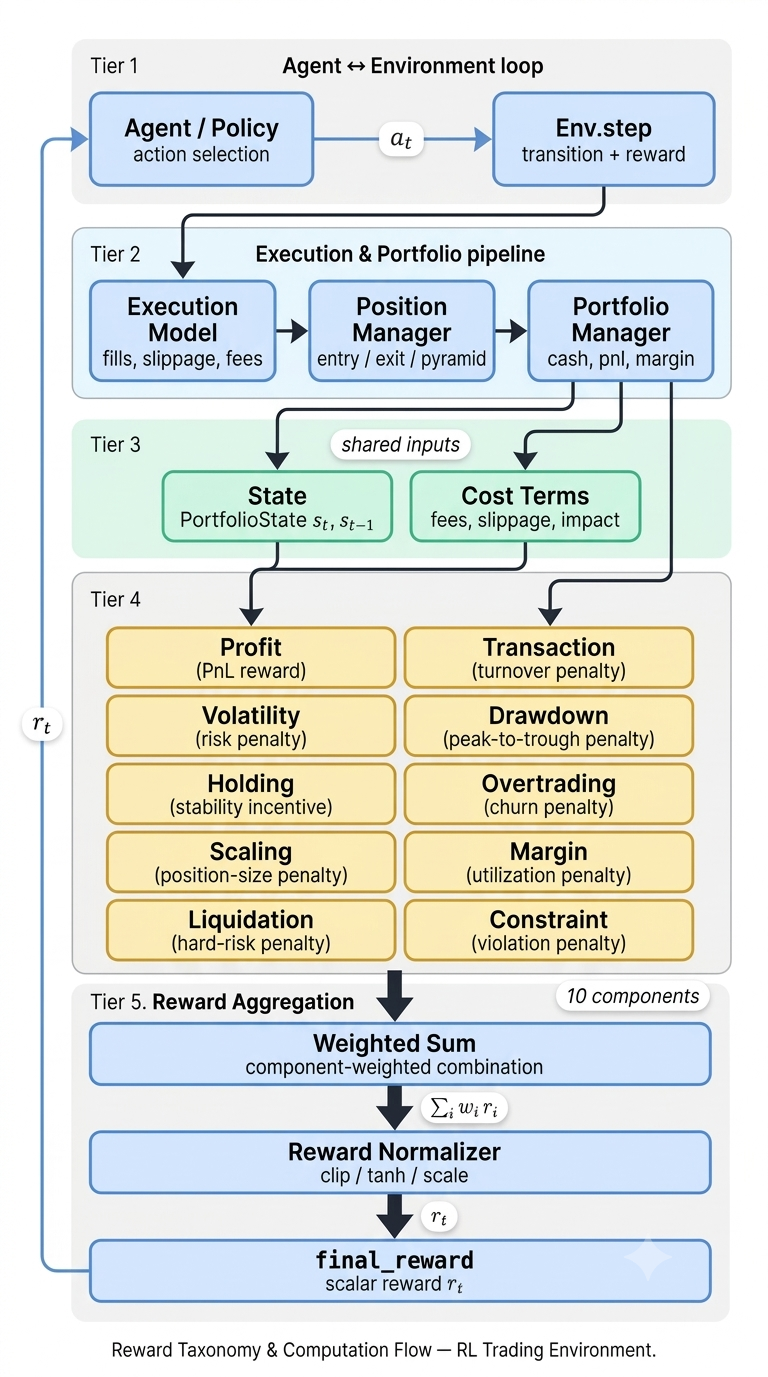}
\caption{Reward taxonomy and computation flow for the RL trading environment, from agent--environment interaction through component aggregation and normalization to the final scalar reward.}
\label{fig:reward_taxonomy}
\end{figure}

The reward algorithm defines the deterministic reward-assembly pipeline: it applies component enable switches, a fixed component order, per-component weighting, clipping, and structured logging. The fixed order is the operational counterpart of the taxonomy summarized in Figure~\ref{fig:reward_taxonomy}.

\begin{algorithm}[!tbp]
\caption{Per-step compositional reward computation.}
\label{alg:reward_computation}
\begin{algorithmic}[1]
\STATE \textbf{Input:} transition trace $\tau_t$, component enable switches $g_i \in \{0,1\}$,
\STATE \hspace*{1em} weights $w_i$, clipping bounds $[r_{\min},r_{\max}]$
\STATE \textbf{Output:} clipped reward $r_t$ and component log dictionary $\mathcal{L}_t$
\STATE Define fixed component order:
\STATE \hspace{1em}profit, holding, volatility, drawdown, transaction, overtrading,
\STATE \hspace{1em}pyramiding, martingale, margin-utilization, liquidation, constraint-violation
\FOR{each component $i$ in the fixed order}
    \IF{$g_i=1$}
        \STATE Compute component value $c_i(\tau_t)$
    \ELSE
        \STATE Set $c_i(\tau_t) \leftarrow 0$
    \ENDIF
    \STATE Compute weighted term $u_i \leftarrow w_i\,c_i(\tau_t)$
    \STATE Append $(c_i,w_i,u_i,g_i)$ to $\mathcal{L}_t$
\ENDFOR
\STATE Compute raw reward $r_t^{\mathrm{raw}} \leftarrow \sum_i u_i$
\STATE Compute clipped reward $r_t \leftarrow \mathrm{clip}(r_t^{\mathrm{raw}}, r_{\min}, r_{\max})$
\STATE Append $(r_t^{\mathrm{raw}}, r_t, \mathbb{1}[r_t^{\mathrm{raw}} \neq r_t])$ to $\mathcal{L}_t$
\STATE Return $(r_t,\mathcal{L}_t)$
\end{algorithmic}
\end{algorithm}

The component dictionary is logged for every transition, while only the clipped scalar reward $r_t$ is passed to the learner. The component traces support diagnosis and ablation attribution without changing the signal consumed by the value update.

\paragraph{Logging and ablation protocol.}
Component evaluation follows a fixed order, and each term is logged per step via the environment \texttt{info} dictionary and persisted reward traces. This enables direct attribution in ablation experiments (\variant{r1}--\variant{r7}), where individual components or weights are selectively modified. Because non-potential reward shaping can alter optimal policies \citep{ng1999policy}, profit remains the primary signal, with auxiliary penalties introduced incrementally.

Figure~\ref{fig:reward_taxonomy} should therefore be read alongside Algorithm~\ref{alg:reward_computation}: it visualizes the same fixed component order and aggregation flow.
\subsection{Agent Architectures}
Both agents are value-based and share the same backbone: a feed-forward multilayer perceptron (MLP) with hidden dimensions $[512,512,256]$, ReLU activations, and a linear Q-value output head. Output dimensionality equals the active action-space size ($3$ in simplified mode, $10$ in extended mode). The agents are Deep Q-Network (DQN) and Double Deep Q-Network (DDQN) variants, selected because they support explicit legality masking in discrete control.

\subsubsection{DQN}
DQN uses temporal-difference targets with periodic hard target-network synchronization \citep{mnih2015humanlevel}:
\begin{equation}
y_t^{\text{DQN}} = r_t + \gamma (1-d_t)\max_{a' \in \mathcal{A}_{\mathrm{legal}}(s_{t+1})} Q_{\theta^-}(s_{t+1}, a').
\label{eq:dqn_target}
\end{equation}
where $y_t^{\text{DQN}}$ is the temporal-difference target, $r_t$ is the clipped reward, $\gamma$ is the discount factor, $d_t \in \{0,1\}$ is the terminal indicator, $\mathcal{A}_{\mathrm{legal}}(s_{t+1})$ is the legal-action set in state $s_{t+1}$, and $Q_{\theta^-}$ is the target-network Q-function parameterized by $\theta^-$. Equation~\ref{eq:dqn_target} masks invalid actions directly in target construction.

\subsubsection{Double DQN}
Double DQN decouples action selection and evaluation:
\begin{equation}
a^* = \arg\max_{a' \in \mathcal{A}_{\mathrm{legal}}(s_{t+1})} Q_{\theta}(s_{t+1}, a').
\label{eq:ddqn_action_select}
\end{equation}
where $a^*$ is the legal greedy action selected by the online network, $Q_{\theta}$ is the online Q-function, $\theta$ denotes the online-network parameters, and $\mathcal{A}_{\mathrm{legal}}(s_{t+1})$ is the legal-action set in state $s_{t+1}$.
\begin{equation}
y_t^{\text{DDQN}} = r_t + \gamma (1-d_t) Q_{\theta^-}(s_{t+1}, a^*).
\label{eq:ddqn_target}
\end{equation}
where $y_t^{\text{DDQN}}$ is the temporal-difference target, $r_t$ is the clipped reward, $\gamma$ is the discount factor, $d_t \in \{0,1\}$ is the terminal indicator, and $Q_{\theta^-}$ is the target-network evaluation at the selected action $a^*$. Equations~\ref{eq:ddqn_action_select} and \ref{eq:ddqn_target} separate online action selection from target-network evaluation, reducing value overestimation in volatile training regimes \citep{hasselt2016deep}.

\paragraph{Mask-aware exploration policy.}
Exploration follows an $\epsilon$-greedy policy defined on legal actions only: with probability $\epsilon$, the agent samples uniformly from legal actions; otherwise, it selects the legal argmax action. The same legality constraints are applied during replay-based target computation, ensuring consistency between data collection and value updates.

\paragraph{Algorithmic variants outside release scope.}
The architecture is compatible with dueling heads and distributional value estimation, which are relevant in heavy-tailed return settings \citep{wang2016dueling,bellemare2017distributional}. These variants are intentionally excluded from the main release protocol to isolate the effect of environment and reward design under DQN-family baselines.

\subsection{Training Protocol}
Core settings are fixed across experiment families unless explicitly overridden. For global reproducibility and side-by-side comparability, the exact training and environment values are consolidated in \Cref{tab:hyperparams} within \Cref{sec:experiments}.

The training algorithm consolidates the mask-aware DQN/DDQN loop, including replay storage, learning gates, target synchronization, and evaluation cadence. In particular, the legal mask $m_t$ is stored in replay and reused during target construction, so exploration and backup share the same legality contract.

\begin{algorithm}[!tbp]
\caption{Mask-aware DQN/DDQN training with legal-action masking.}
\label{alg:mask_aware_training}
\begin{algorithmic}[1]
\STATE \textbf{Input:}
\STATE \hspace*{1em} \textbf{Environment \,\& state:} environment $\mathcal{E}$; state $s_t$; legal-action mask $m_t \in \{0,1\}^{|\mathcal{A}|}$, where $m_t[a]=1$ iff $a$ is legal in $s_t$
\STATE \hspace*{1em} \textbf{Networks:} online network $Q_{\theta}$; target network $Q_{\theta^-}$
\STATE \hspace*{1em} \textbf{Replay buffer:} $\mathcal{D}$ with capacity $N$
\STATE \hspace*{1em} \textbf{Training hyperparameters:} batch size $B$; discount $\gamma$
\STATE \hspace*{1em} \textbf{Scheduling:} learn-start $t_{\mathrm{start}}$; learn frequency $f_{\mathrm{learn}}$; target-sync interval $f_{\mathrm{target}}$; evaluation interval $f_{\mathrm{eval}}$; total steps $T$; evaluation budget $K_{\mathrm{eval}}$
\STATE \hspace*{1em} \textbf{Algorithm type:} $\texttt{DQN}$ or $\texttt{DDQN}$
\STATE Initialize $\theta$ randomly; set $\theta^- \leftarrow \theta$; initialize the $\epsilon$ schedule
\STATE Reset $\mathcal{E}$ and observe $(s_0,m_0)$
\FOR{$t=0$ to $T-1$}
    \STATE Select $a_t$ by mask-aware $\epsilon$-greedy on $m_t$: with probability $\epsilon$, sample $a_t \sim \mathrm{Unif}(\{a:m_t[a]=1\})$; otherwise $a_t \leftarrow \arg\max_{a:\,m_t[a]=1} Q_{\theta}(s_t,a)$
    \STATE Execute $a_t$ in $\mathcal{E}$ and observe $(s'_{t},r_t,d_t,\mathrm{info}_t,m'_{t})$
    \STATE Store $(s_t,a_t,r_t,s'_{t},d_t,m_t,m'_{t})$ in $\mathcal{D}$
    \STATE Update $(s_t,m_t) \leftarrow (s'_{t},m'_{t})$
    \IF{$t \ge t_{\mathrm{start}}$ \textbf{and} $t \bmod f_{\mathrm{learn}}=0$ \textbf{and} $|\mathcal{D}| \ge B$}
        \STATE Sample minibatch $\{(s_i,a_i,r_i,s'_i,d_i,m_i,m'_i)\}_{i=1}^{B}$ from $\mathcal{D}$
        \IF{algorithm type is \texttt{DDQN}}
            \STATE $a_i^{\star} \leftarrow \arg\max_{a:\,m'_i[a]=1} Q_{\theta}(s'_i,a),\ \forall i$
            \STATE $y_i \leftarrow r_i + \gamma(1-d_i)Q_{\theta^-}(s'_i,a_i^{\star}),\ \forall i$
        \ELSE
            \STATE $y_i \leftarrow r_i + \gamma(1-d_i)\max_{a:\,m'_i[a]=1} Q_{\theta^-}(s'_i,a),\ \forall i$
        \ENDIF
        \STATE $\mathcal{L} \leftarrow \frac{1}{B}\sum_{i=1}^{B}\mathrm{Huber}\big(Q_{\theta}(s_i,a_i)-y_i\big)$
        \STATE Update $\theta$ by gradient descent with gradient clipping
    \ENDIF
    \IF{$t \bmod f_{\mathrm{target}}=0$}
        \STATE Hard-sync target network: $\theta^- \leftarrow \theta$
    \ENDIF
    \IF{$t \bmod f_{\mathrm{eval}}=0$}
        \STATE Evaluate the deterministic policy ($\epsilon=0$) for $K_{\mathrm{eval}}$ episodes and log all outputs
    \ENDIF
    \STATE Update $\epsilon$
\ENDFOR
\STATE \textbf{Output:} trained parameters $\theta$, replay buffer state, and diagnostic artifacts
\end{algorithmic}
\end{algorithm}

Algorithm~\ref{alg:mask_aware_training} consumes the transition and reward outputs described above and returns trained parameters together with diagnostic artifacts. The subsequent figure is a schematic of this training algorithm, not a separate procedure.
\begin{figure}[tbp]
\centering
\includegraphics[width=\linewidth,height=0.8\textheight,keepaspectratio]{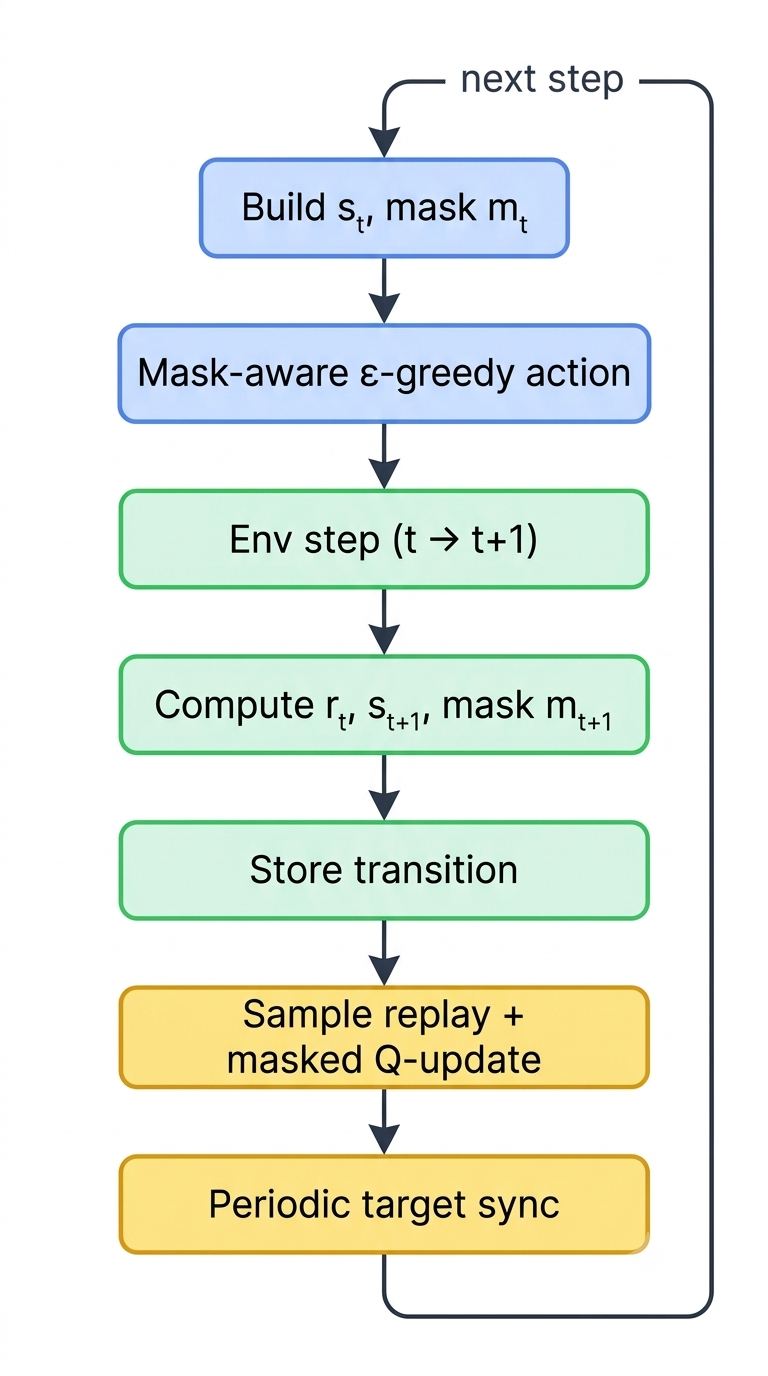}
\caption{Training-loop schematic with mask-aware interaction, complete transition construction, replay learning, and periodic target synchronization.}
\label{fig:training_loop}
\end{figure}

\paragraph{Rationale for optimization settings.}
The $10^6$-step horizon improves coverage of diverse market regimes for off-policy replay. A 40K replay buffer balances decorrelation and recency under non-stationary financial dynamics. Batch size 128, Huber loss, gradient clipping (norm 10), and a learning rate of $2.5\times10^{-4}$ were selected to stabilize updates without obscuring the effect of the environment and reward design. Detailed values are summarized in Table~\ref{tab:hyperparams} and Appendix~\ref{app:hyperparams}.

\paragraph{Evaluation cadence and diagnostic logging.}
Training logs include learning curves, loss, reward components, action distributions, turnover, and liquidation events. Periodic deterministic evaluation is performed during training (default interval: every 10,000 steps) using training episodes and stored trajectories. At the end of each run, a deterministic full-horizon evaluation is conducted on the training data. Reported metrics include cumulative and annualized returns, volatility, Sharpe~\citep{sharpe1994sharpe}, Sortino~\citep{sortino1994performance}, maximum drawdown, win rate, turnover, liquidation statistics, and scaling utilization. These evaluations follow the scope stated above.

\paragraph{Determinism scope and methodological limitation.}
The canonical seed and the remaining reproducibility details are reported in Section~\ref{sec:experiments}.

\section{Experimental Design}
\label{sec:experiments}

This section specifies the controlled evaluation protocol used to answer \RQ{1}--\RQ{3}. We first define scope and invariants, then map each experiment family to a research question, and finally describe metrics and reproducibility controls.

\subsection{Dataset Scope and Split Policy}
The primary controlled families (\Exp{01}--\Exp{03}) are executed on hourly EURUSD training-split data with Double DQN~\citep{hasselt2016deep} to isolate component effects under a fixed optimization budget. We additionally report benchmark comparisons and cross-pair diagnostics from the same artifact generation pipeline to contextualize the primary findings; these are descriptive analyses rather than transfer benchmarks.

\begin{table}[!tbp]
\centering
\caption{Dataset statistics (raw hourly bars). The chronological 80/20 split is
shown for completeness; all reported experiments use the training split only.}
\label{tab:dataset}
\small
\setlength{\tabcolsep}{6pt}

\begin{tabular}{@{}lrrr@{}}
\toprule
\textbf{Pair} & \textbf{$N_{\text{total}}$} & \textbf{$N_{\text{train}}$} & \textbf{$N_{\text{held-out}}$} \\
\midrule
AUDUSD & 25\,000 & 20\,000 & 5\,000 \\
EURUSD & 25\,000 & 20\,000 & 5\,000 \\
GBPUSD & 25\,000 & 20\,000 & 5\,000 \\
USDJPY & 25\,000 & 20\,000 & 5\,000 \\
\bottomrule
\end{tabular}

\vspace{1.2ex}

\adjustbox{max width=\linewidth}{%
\begin{tabular}{@{}lcc@{}}
\toprule
\textbf{Pair} & \textbf{Train period} & \textbf{Held-out period$^{\dagger}$} \\
\midrule
AUDUSD & 2022-01-01 -- 2025-01-01 & 2025-01-01 -- 2026-01-01 \\
EURUSD & 2022-01-01 -- 2025-01-01 & 2025-01-01 -- 2026-01-01 \\
GBPUSD & 2022-01-01 -- 2025-01-01 & 2025-01-01 -- 2026-01-01 \\
USDJPY & 2022-01-01 -- 2025-01-01 & 2025-01-01 -- 2026-01-01 \\
\bottomrule
\end{tabular}}

\vspace{0.4ex}
{\footnotesize $^{\dagger}$Held-out period reported for completeness only;
not used in any reported experiment.}
\end{table}

\noindent The dataset table clarifies that all pairs share an aligned chronological split policy, allowing pair-level comparisons without confounding differences in temporal coverage.

\subsection{Experiment Families and Contextual Extensions}
All studies are executed via configuration-only overrides using YAML files, with no structural code changes:
\begin{itemize}[leftmargin=1.4em]
	\item \textbf{\Exp{01} Reward ablation}: cumulative reward component schedule \variant{r1} to \variant{r7}.
	\item \textbf{\Exp{02} Action space}: simplified (3-action adapter) vs extended (10-action).
	\item \textbf{\Exp{03} Scaling analysis}: no scaling, pyramid-only, martingale-only, both.
    \item \textbf{Context A (benchmark calibration)}: RL agents compared against random, buy-and-hold, momentum, and mean-reversion baselines under identical simulator semantics.
	\item \textbf{Context B (cross-pair diagnostics)}: DQN vs. Double DQN comparisons on GBPUSD, USDJPY, and AUDUSD using the full reward profile; these are descriptive observations under the same simulator semantics.
\end{itemize}
All families inherit the same anti-lookahead timing rule, transaction-friction model, and margin constraints, ensuring that observed differences arise from controlled design changes rather than simulator drift.

\subsection{Protocol Invariants and Hyperparameter Budget}
To preserve comparability, core optimization and environment settings are held constant across families unless a specific family intentionally perturbs that axis. These shared values are summarized in \Cref{tab:hyperparams}.

\begin{table*}[!tbp]
\centering
\caption{Core hyperparameters, interface dimensions, and environment settings used in release experiments.}
\label{tab:hyperparams}

\begin{adjustbox}{width=\linewidth}
\begin{tabularx}{\linewidth}{@{}lXl@{}}
\toprule
\textbf{Parameter} & \textbf{Value} & \textbf{Scope} \\
\midrule
Observation schema & Dict:\;market/portfolio/mask/flat & Environment \\
Observation window & 24 bars & Environment \\
Portfolio vector dimension & 10 & Environment \\
Action outputs ($n_a$) & 10 (extended) or 3 (simplified) & Agent/Environment \\
Flat input dimension & $24\,d_{\mathrm{feat}} + 10 + n_a$ & Agent \\
Q-network architecture & MLP [512, 512, 256] + ReLU + linear head & Agent \\
Total timesteps & 1,000,000 & Training \\
Replay buffer size & 40,000 & Training \\
Batch size & 128 & Training \\
Learn start & 10,000 steps & Training \\
Learn frequency & every 4 steps & Training \\
$\gamma$ & 0.99 & Training \\
Optimizer & Adam (2.5e-4) & Agent \\
Epsilon schedule & 1.0$\rightarrow$0.01 (30K) & Agent \\
Target update interval & 2,000 steps & Agent \\
Gradient clip & 10.0 & Agent \\
Periodic diagnostic evaluation & every 10,000 steps (default) & Training \\
Automatic mixed precision (AMP) dtype & fp16 & Agent \\
Initial capital & USD 100,000 & Environment \\
Max leverage & 30x & Environment \\
Liquidation threshold & 25\% equity & Environment \\
Commission & USD 3.5 / lot (round trip) & Environment \\
Base slippage & 0.5 pips & Environment \\
Seed & See Section~\ref{sec:experiments} & Reproducibility \\
Temporal split & 80/20 chronological & Reproducibility \\
\bottomrule
\end{tabularx}
\end{adjustbox}
\end{table*}

\noindent Fixing this budget is especially important when interpreting action-space and scaling effects, because it prevents confounding from unequal replay exposure or optimizer aggressiveness.

\subsection{Research Questions}
Each experiment family is explicitly tied to a research question that guides design choices and governs interpretation of outcomes:
\begin{description}[leftmargin=1.8em, style=nextline]
    \item[\textbf{RQ1} (Reward Interactions):] Do reward components interact non-monotonically, and does progressive penalty activation improve or degrade training-period risk-adjusted return compared to a bare profit signal? (\Exp{01})
	\item[\textbf{RQ2} (Action Granularity):] How does the extended 10-action interface change return, risk-adjusted performance, drawdown, and turnover relative to the 3-action coarse adapter? (\Exp{02})
	\item[\textbf{RQ3} (Scaling Asymmetry):] How does the $2.4\times$ martingale-to-pyramid penalty ratio relate to asymmetric position-management behavior, with pyramid-only variants outperforming martingale-only on drawdown? (\Exp{03})
\end{description}
Context A and Context B are interpreted as calibration analyses rather than additional primary RQs; they are used to situate the main EURUSD findings against stronger baselines and pair variation.

\subsection{Metrics and Statistical Testing}
Evaluation captures profitability, risk, and behavior: cumulative return, annualized return/volatility, Sharpe~\citep{sharpe1994sharpe}, Sortino~\citep{sortino1994performance}, maximum drawdown, win rate, turnover, liquidation metrics, and scaling utilization. For paired model comparisons, outcomes are reported descriptively; no inferential hypothesis testing is performed.

Because all reported values are endpoint artifacts, we emphasize effect patterns and relative trade-offs over formal claims of distributional superiority. See Section~\ref{sec:experiments} for canonical seeding and evaluation scope, consistent with cautionary guidance in trading-RL evaluation practice \citep{zhang2020deep}.

\subsection{Reproducibility Protocol}
Reproducibility controls include deterministic seeding across Python/NumPy/PyTorch/environment, resolved-config snapshots stored per run, canonical artifact paths, and strict chronological splitting. Numerical findings in this manuscript are therefore directly traceable to saved outputs. This protocol is aligned with reproducibility priorities emphasized in prior trading-RL benchmark literature~\citep{liu2021finrl,zhang2020deep}.

\paragraph{Important scope note.}
This release reports \emph{single-seed} outcomes (seed $42$). Accordingly, we interpret differences as informative system-behavior evidence rather than definitive estimates of expected out-of-sample superiority.

\section{Results}
\label{sec:results}

All values in this section are derived from \emph{training artifacts only} (see Section~\ref{sec:experiments}). We report primary outcomes for \Exp{01}--\Exp{03}, then place them in context using benchmark comparisons and cross-pair diagnostics generated under the same simulator semantics. These latter analyses are descriptive and do not establish transfer performance.

\subsection{Global Performance Dashboard}

\begin{figure}[!tbp]
\centering
\includegraphics[width=\linewidth]{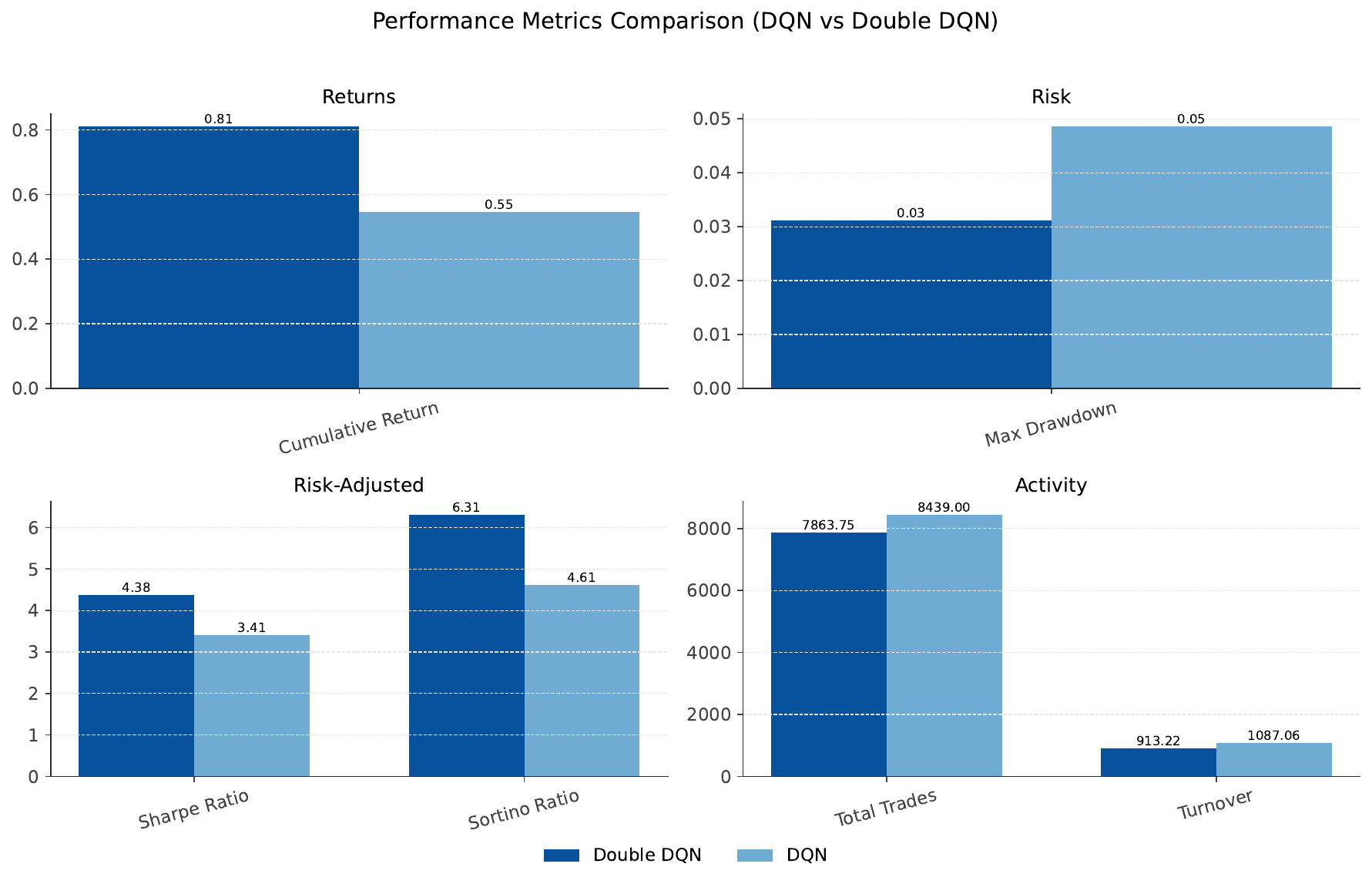}
\caption{Multi-panel dashboard of endpoint metrics across retained experiment families, summarizing return, drawdown, risk-adjusted performance, and activity.}
\label{fig:metrics_multi_panel}
\end{figure}

Before unpacking each experiment family, \Cref{fig:metrics_multi_panel} provides a compact multi-metric snapshot of profitability, risk, and behavior indicators used throughout this section.

The dashboard suggests that no single configuration dominates all dimensions simultaneously: variants with stronger endpoint return often incur higher activity or weaker conservative risk summaries, reinforcing the need for objective-specific model selection.

\subsection{Reward Component Ablation (\Exp{01})}

\noindent\textbf{Objective.}
\Exp{01} progressively enables reward components to identify constructive versus destructive interactions.

We first report endpoint statistics in \Cref{tab:ablation_train}, then use distributional and trajectory views to diagnose interaction nonlinearity.

\begin{table*}[!tbp]
\centering
\caption{Reward ablation on EURUSD train split (Double DQN; see Section~\ref{sec:experiments}).}
\label{tab:ablation_train}

\begin{adjustbox}{width=\linewidth}
\begin{tabular}{lrrrrrrrr}
\toprule
\textbf{Variant} & \textbf{Sharpe} & \textbf{Sortino} & \textbf{Cum.Ret(\%)} & \textbf{MaxDD(\%)} & \textbf{WinRate(\%)} & \textbf{Turnover} & \textbf{AvgPyr} & \textbf{AvgMart} \\
\midrule
\variant{r1} & 0.412 & 0.573 & 27.79 & 10.48 & 35.18 & 1592.83 & 0.113 & 0.121 \\
\variant{r2} & 0.237 & 0.353 & 12.20 & 8.60 & 31.85 & 1741.58 & 0.187 & 0.127 \\
\variant{r3} & 0.638 & 0.979 & 41.20 & 5.44 & 31.53 & 1488.67 & 0.165 & 0.138 \\
\variant{r4} & 0.687 & 1.029 & 43.21 & 5.68 & 30.99 & 1344.47 & 0.197 & 0.144 \\
\variant{r5} & 0.589 & 0.822 & 41.74 & 4.12 & 31.07 & 1350.00 & 0.172 & 0.156 \\
\variant{r6} & 0.410 & 0.539 & 24.58 & 7.70 & 27.38 & 1200.32 & 0.132 & 0.089 \\
\variant{r7} & 0.765 & 1.117 & 57.09 & 2.31 & 33.15 & 1156.51 & 0.173 & 0.169 \\
\bottomrule
\end{tabular}
\end{adjustbox}
\end{table*}

Endpoint values already indicate interaction effects: Sharpe and cumulative return do not improve monotonically with added penalties, and several intermediate variants underperform simpler compositions.

\begin{figure}[!tbp]
\centering
\includegraphics[width=\linewidth]{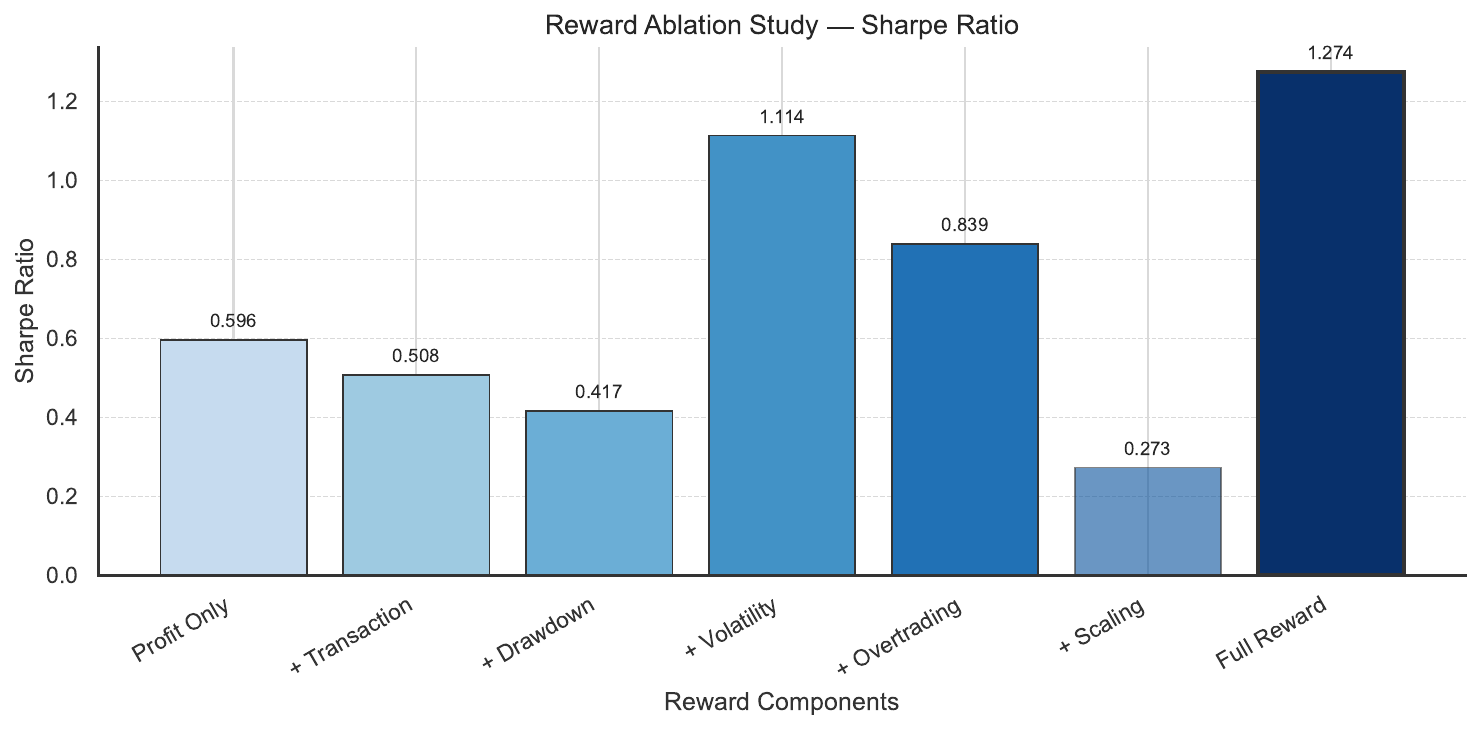}
\caption{Reward-ablation comparison for variants \variant{r1}--\variant{r7} on EURUSD, showing a non-monotone profile with \variant{r7} as the strongest endpoint.}
\label{fig:ablation_bar}
\end{figure}

\Cref{fig:ablation_bar} makes the non-monotonicity explicit: the sequence deteriorates from \variant{r1} to \variant{r2}, recovers across \variant{r3}--\variant{r5}, softens at \variant{r6}, and peaks at \variant{r7}. The pattern is consistent with interaction effects among reward components rather than a strictly additive response to additional penalties.

\begin{figure}[!tbp]
\centering
\includegraphics[width=\linewidth]{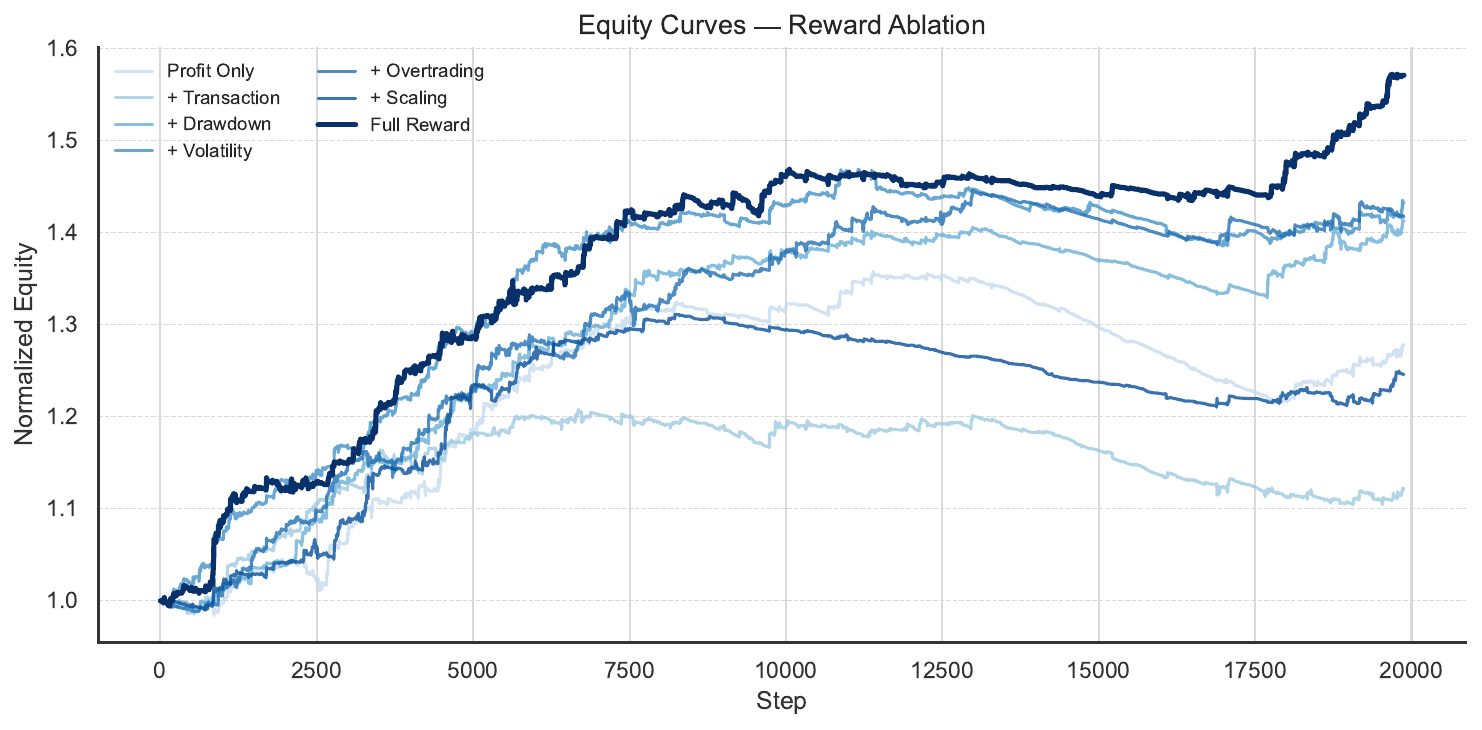}
\caption{Training equity trajectories for \variant{r1}--\variant{r7}, highlighting non-monotone intermediate behavior and the strongest endpoint for \variant{r7}.}
\label{fig:equity_curves_ablation}
\end{figure}

The trajectory view in \Cref{fig:equity_curves_ablation} corroborates the endpoint summary: the strongest terminal profile is not produced by the earliest penalty additions, but by the fully calibrated composition. This supports the claim that reward engineering is an empirical calibration task rather than a strictly additive design process.

\subsection{Action-Space Comparison (\Exp{02})}

\noindent\textbf{Objective.}
\Exp{02} compares the extended 10-action interface with the simplified 3-action adapter under matched training settings.

\begin{table}[!tbp]
\centering
\caption{Action-space comparison on the EURUSD (Double DQN).}
\label{tab:action_space}
\begin{adjustbox}{width=\linewidth}
\begin{tabular}{@{}lrrrrr@{}}
\toprule
\textbf{Action Space} & \textbf{Sharpe} & \textbf{Cum.Ret(\%)} & \textbf{MaxDD(\%)} & \textbf{WinRate(\%)} & \textbf{Turnover} \\
\midrule
Extended (10) & 0.765 & 57.09 & 2.31 & 33.15 & 1156.51 \\
Simplified (3) & 2.433 & 33.21 & 0.29 & 68.13 & 528.65  \\
\bottomrule
\end{tabular}
\end{adjustbox}
\end{table}

As shown in \Cref{tab:action_space}, the extended interface improves cumulative return while increasing turnover and reducing Sharpe relative to the conservative adapter, indicating a return--activity trade-off under the fixed training budget.

\begin{figure}[!tbp]
\centering
\includegraphics[width=\linewidth]{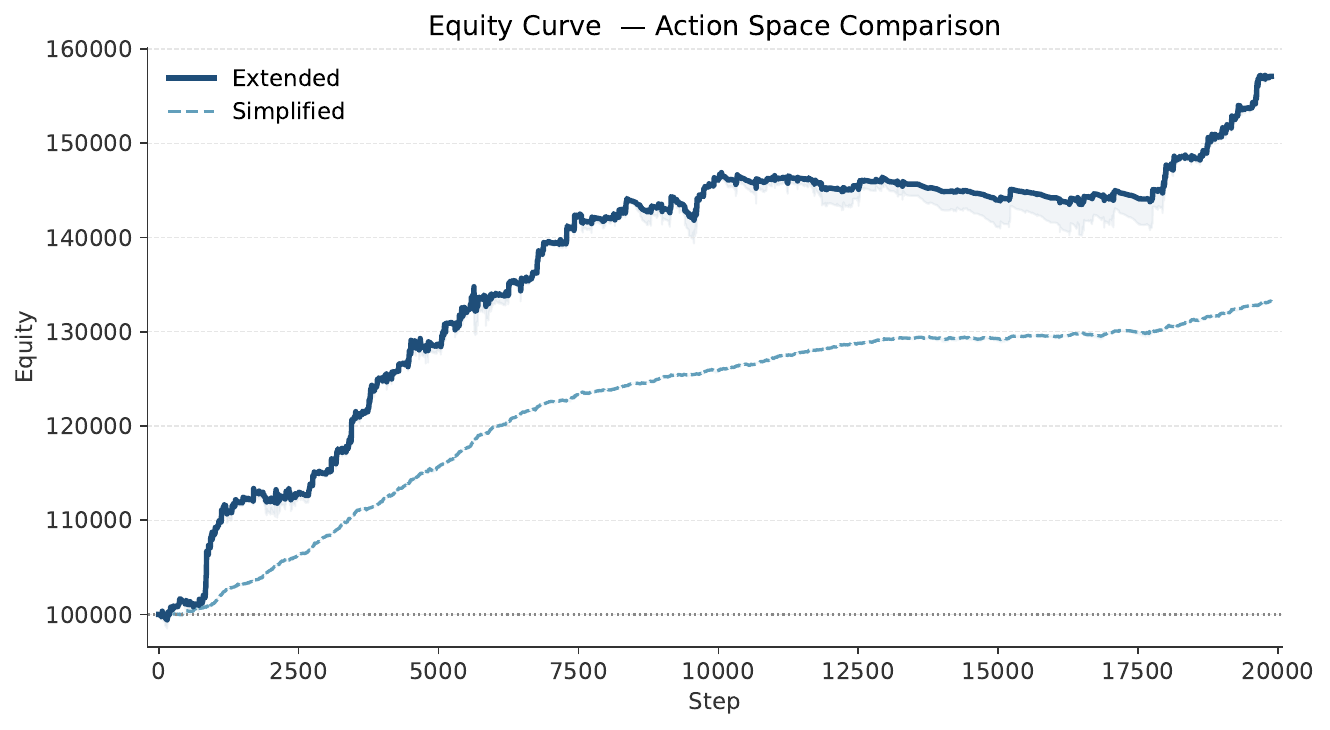}
\caption{Equity-curve comparison of simplified (3-action) and extended (10-action) interfaces on EURUSD, showing a return--smoothness trade-off.}
\label{fig:action_space_equity_comparison}
\end{figure}

To visualize how this trade-off accumulates over the episode rather than only at endpoint, \Cref{fig:action_space_equity_comparison} contrasts the corresponding equity paths.

The temporal separation in \Cref{fig:action_space_equity_comparison} clarifies that action granularity changes policy behavior throughout training, not only final-point metrics: richer primitives increase opportunity capture but amplify path variability.

\subsection{Scaling Strategy Analysis (\Exp{03})}

\noindent\textbf{Objective.}
\Exp{03} evaluates four scaling configurations: no scaling (\variant{s1}), pyramid-only (\variant{s2}), martingale-only (\variant{s3}), and combined (\variant{s4}).

\begin{table}[!tbp]
\centering
\caption{Scaling-strategy analysis on the EURUSD (Double DQN).}
\label{tab:scaling}
\begin{adjustbox}{width=\linewidth}
\begin{tabular}{@{}lrrrrrr@{}}
\toprule
\textbf{Variant} & \textbf{Sharpe} & \textbf{Cum.Ret(\%)} & \textbf{MaxDD(\%)} & \textbf{AvgPyr} & \textbf{AvgMart} & \textbf{Turnover} \\
\midrule
s1\_no\_scaling & -1.568 & -16.54 & 16.74 & 0.000 & 0.000 & 1440.09 \\
s2\_pyramid & -0.136 & -2.48 & 5.73 & 0.083 & 0.000 & 1394.52 \\
s3\_martingale & 0.319 & 8.94 & 4.63 & 0.000 & 0.122 & 1273.71 \\
s4\_both & 0.355 & 13.08 & 6.79 & 0.086 & 0.096 & 1483.55 \\
\bottomrule
\end{tabular}
\end{adjustbox}
\end{table}

\Cref{tab:scaling} shows that all scaling-enabled variants improve over no scaling on drawdown, with distinct objective-dependent preferences across return and drawdown.

\begin{figure}[!tbp]
\centering
\includegraphics[width=\linewidth]{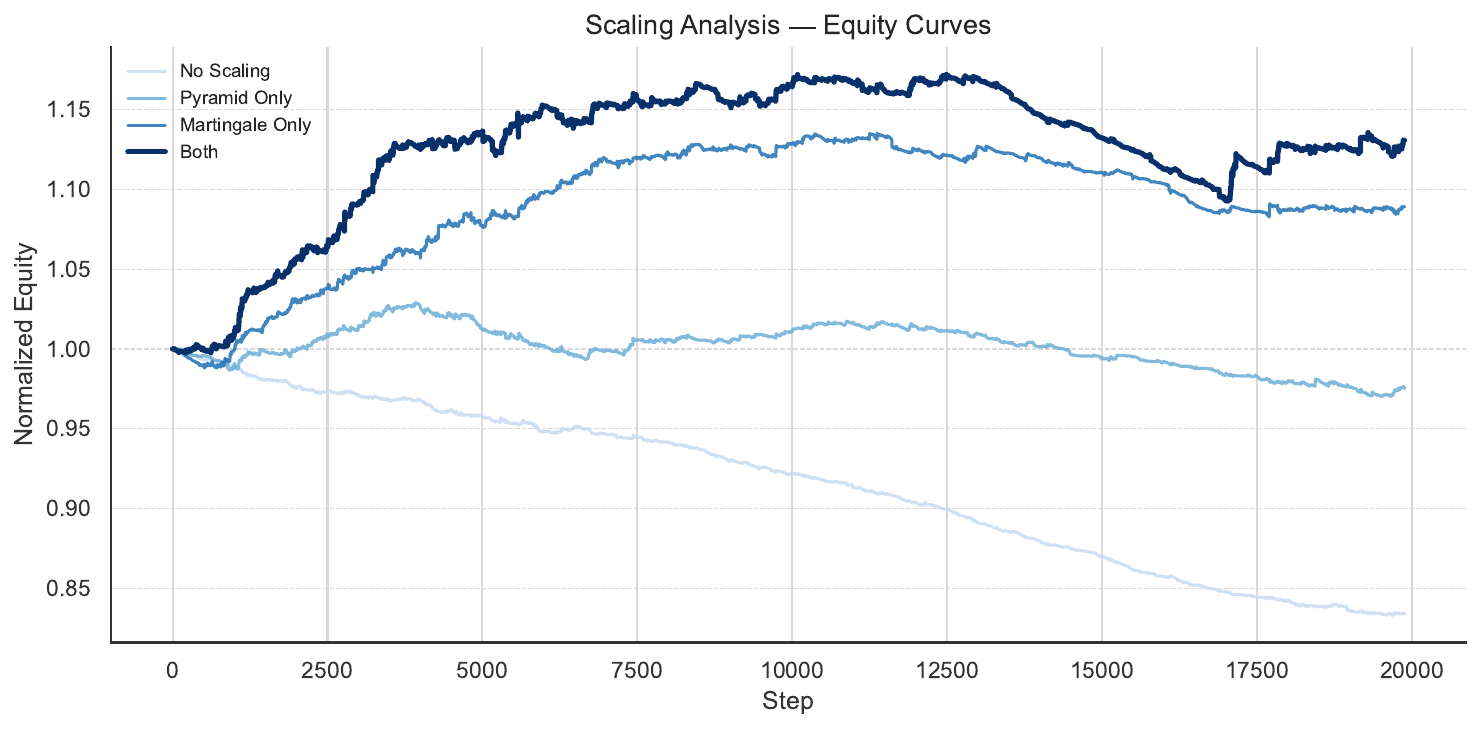}
\caption{Training equity trajectories for scaling variants \variant{s1}--\variant{s4} on EURUSD, where scaling-enabled variants reduce maximum drawdown versus no scaling.}
\label{fig:equity_curves_scaling}
\end{figure}

The path-level evidence in \Cref{fig:equity_curves_scaling} suggests that enabling scaling alters drawdown dynamics over the full training horizon rather than merely shifting a single endpoint.

\subsection{Benchmark Calibration on EURUSD}

Primary RQ findings are further calibrated against non-RL baselines and a DQN comparator. We first report headline endpoint metrics in \Cref{tab:benchmarks}, then inspect complementary risk diagnostics in \Cref{tab:benchmark_risk}.

\begin{table}[!tbp]
\centering
\caption{Benchmark comparison on the EURUSD training split. RL agents use full reward variant \variant{r7}.}
\label{tab:benchmarks}
\begin{adjustbox}{width=\linewidth}
\begin{tabular}{@{}lrrrrr@{}}
\toprule
\textbf{Method} & \textbf{Sharpe} & \textbf{Cum.Ret(\%)} & \textbf{MaxDD(\%)} & \textbf{WinRate(\%)} & \textbf{Turnover} \\
\midrule
Random Policy & -13.380 & -34.58 & 34.70 & 25.99 & 2199.98 \\
Buy-and-Hold & -0.358 & -1.05 & 2.08 & 0.00 & 0.11 \\
Momentum & -1.355 & -3.82 & 4.05 & 23.62 & 186.28 \\
Mean-Reversion & -0.807 & -1.67 & 2.34 & 24.64 & 157.67 \\
DQN & 0.369 & 24.56 & 8.72 & 34.12 & 1316.51 \\
Double DQN & 0.765 & 57.09 & 2.31 & 33.15 & 1156.51 \\
\bottomrule
\end{tabular}
\end{adjustbox}
\end{table}

\begin{table}[!tbp]
\centering
\caption{Benchmark risk diagnostics on the EURUSD (see Section~\ref{sec:experiments}).}
\label{tab:benchmark_risk}
\begin{adjustbox}{width=\linewidth}
\begin{tabular}{@{}lrrrrr@{}}
\toprule
\textbf{Method} & \textbf{Ann.Ret(\%)} & \textbf{Ann.Vol(\%)} & \textbf{Sortino} & \textbf{Trades} \\
\midrule
Random Policy & -12.11 & 0.96 & -16.523 & 13728 \\
Buy-and-Hold & -0.32 & 0.89 & -0.466 & 1 \\
Momentum & -1.18 & 0.87 & -1.789 & 1715 \\
Mean-Reversion & -0.51 & 0.63 & -0.772 & 1469 \\
DQN & 6.36 & 3.37 & 2.643 & 11240 \\
Double DQN & 14.82 & 4.11 & 4.771 & 8415 \\
\bottomrule
\end{tabular}
\end{adjustbox}
\end{table}

Taken together, the two tables show that RL improvements are not merely profit inflation: Double DQN remains favorable under risk-aware summaries and avoids liquidation events in this artifact view, while rule-based baselines underperform on cumulative return. These comparisons are training-split calibrations, not deployment claims.

\begin{figure}[!tbp]
\centering
\includegraphics[width=\linewidth]{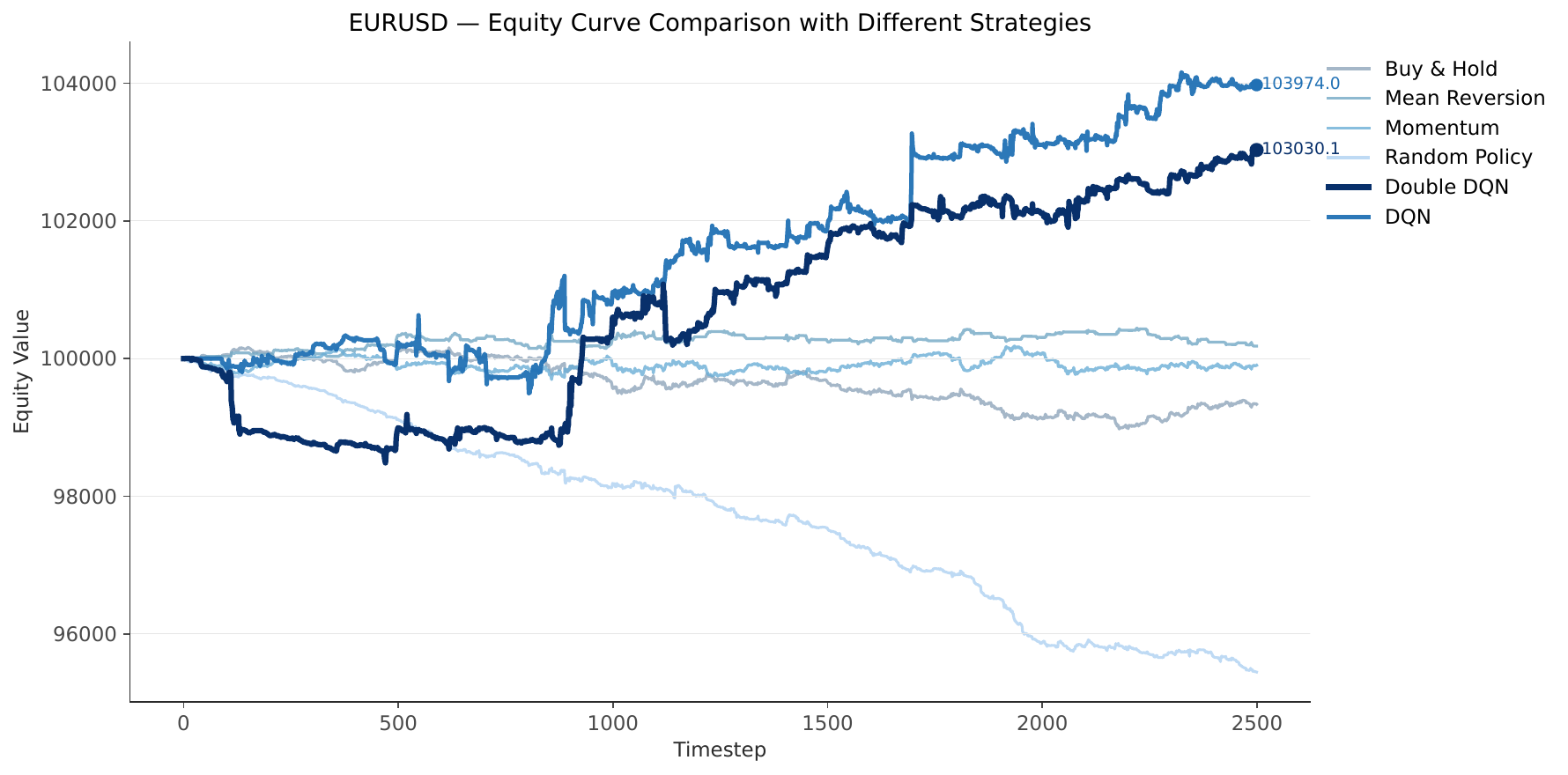}
\caption{Equity trajectories for RL agents and benchmark strategies on EURUSD, with rule-based baselines as calibration anchors.}
\label{fig:benchmark_equity_comparison}
\end{figure}

\Cref{fig:benchmark_equity_comparison} reinforces that the learning-based agents maintain stronger growth profiles than benchmark heuristics under identical execution assumptions.

\subsection{Cross-Pair Generality Diagnostics}

To assess whether the EURUSD pattern is idiosyncratic, \Cref{tab:cross_pair} reports DQN-vs-DDQN outcomes on three additional pairs using the same full-reward configuration.

\begin{table*}[!tbp]
\centering
\caption{Cross-pair DQN vs.\ Double DQN  under the full-reward variant \variant{r7}; results are descriptive diagnostics only.}
\label{tab:cross_pair}

\begin{adjustbox}{width=\linewidth}
\begin{tabular}{llrrrrr}
\toprule
\textbf{Pair} & \textbf{Agent} & \textbf{Sharpe} & \textbf{Sortino} & \textbf{Cum.Ret(\%)} & \textbf{MaxDD(\%)} & \textbf{WinRate(\%)} \\
\midrule
EURUSD & DQN & 0.369 & 0.501 & 24.53 & 8.72 & 34.15 \\
EURUSD & DDQN & 0.765 & 1.117 & 57.09 & 2.31 & 33.15 \\
\midrule
GBPUSD & DQN & 0.164 & 0.195 & 11.98 & 9.59 & 31.96 \\
GBPUSD & DDQN & 0.759 & 0.969 & 83.19 & 3.14 & 31.90 \\
\midrule
USDJPY & DQN & 0.958 & 1.419 & 94.08 & 2.16 & 35.63 \\
USDJPY & DDQN & 1.115 & 1.615 & 115.01 & 4.44 & 33.69 \\
\midrule
AUDUSD & DQN & 0.542 & 0.754 & 33.93 & 7.63 & 30.91 \\
AUDUSD & DDQN & 0.811 & 1.222 & 52.98 & 4.91 & 32.16 \\
\bottomrule
\end{tabular}
\end{adjustbox}
\end{table*}

These results are restricted to the experimental scope defined in Section~\ref{sec:experiments}; the directional trend across pairs is descriptive. Double DQN is generally stronger on return-oriented metrics, but the pair-specific drawdown behavior varies, so the table should not be read as a transfer benchmark.

\begin{figure}[!tbp]
\centering
\includegraphics[width=\linewidth]{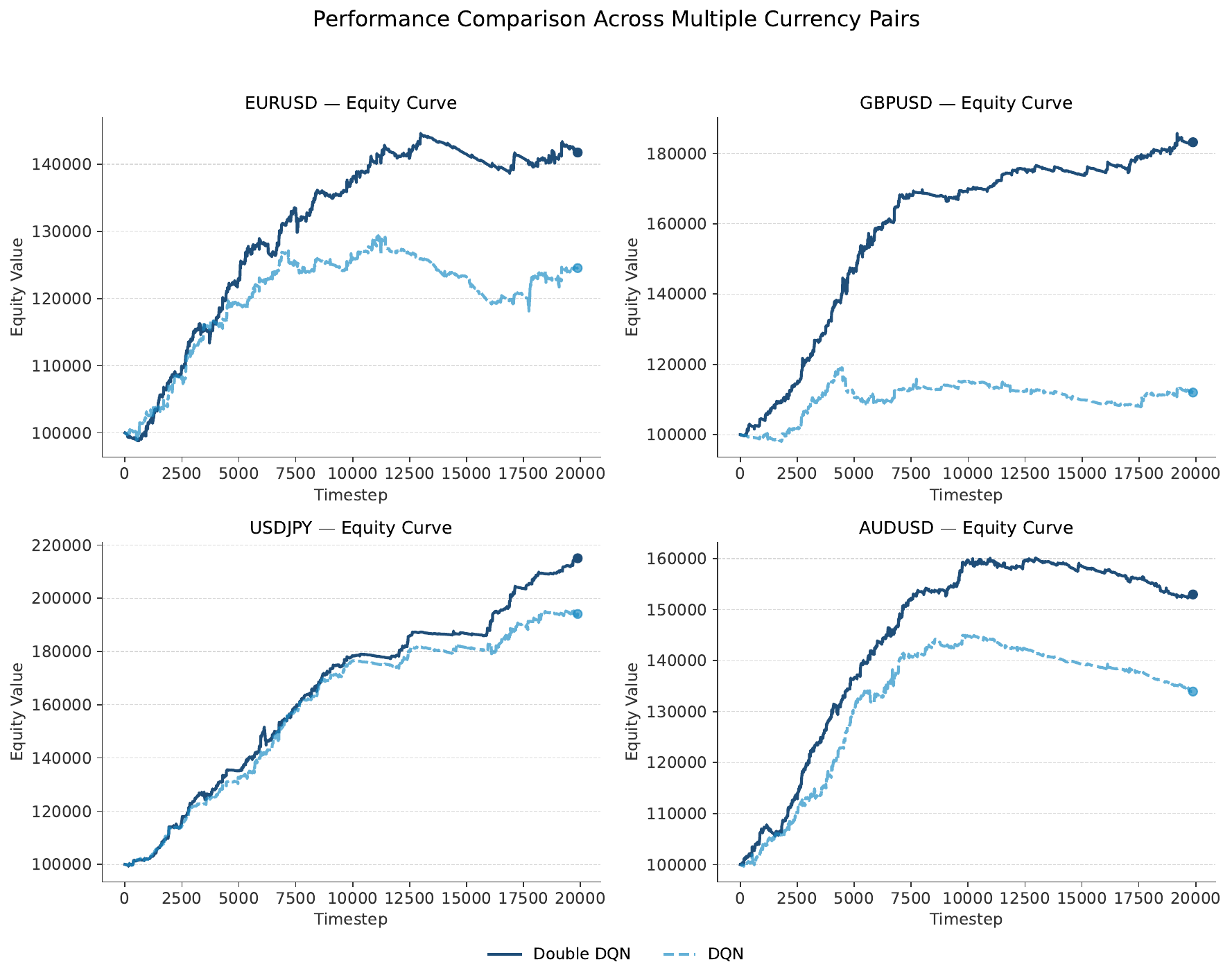}
\caption{Multi-pair equity trajectories (EURUSD, GBPUSD, USDJPY, AUDUSD) for value-based agents under aligned simulator and reward settings.}
\label{fig:multi_pair_equity_curves}
\end{figure}

\Cref{fig:multi_pair_equity_curves} illustrates that cross-pair behavior is heterogeneous in amplitude but consistent in qualitative ordering. This supports the use of pair-diverse diagnostics alongside the primary EURUSD-focused research questions, but it does not establish cross-market generalization.
\section{Discussion}
\label{sec:discussion}

This section interprets the empirical findings through the lens of design objectives introduced earlier: transparent reward construction, legality-aware action control, and causal simulation fidelity. We focus on what can be inferred from controlled experimental evidence (see Section~\ref{sec:experiments}) and explicitly separate methodological insights from broader deployment claims, aligning with the evaluation rigor advocated by \citep{zhang2020deep}.

\subsection{Reward Decomposition as an Experimental Instrument}
The principal methodological insight of this manuscript is that reward engineering should be treated as controlled experimentation rather than iterative tuning. The 11-component architecture, together with deterministic per-step logging, enables direct attribution of behavioral changes to specific reward terms. In \Exp{01}, performance does not improve monotonically as components are added; instead, quality degrades and recovers across the schedule before peaking at \variant{r7}. This pattern demonstrates interaction effects that are difficult to diagnose under monolithic scalar objectives, which are commonly used in trading RL to optimize raw return or equity delta \citep{meng2019reinforcement,theate2021application}. Unlike these conventional approaches, which obscure component-level attribution and complicate reward shaping analysis \citep{ng1999policy}, our decomposable framework isolates the specific penalties that suppress excessive trading and the signals that improve entry timing. The fixed weights are therefore best interpreted as a calibrated baseline: profit remains dominant, but auxiliary terms regularize turnover, drawdown, and leverage usage. The ablation schedule shows that modest changes in component activation can materially shift the return--drawdown balance, which is precisely why the paper treats reward composition as an empirical design problem.

\subsection{Behavioral Consequences of Action Granularity}
\begin{figure}[!tbp]
\centering
\includegraphics[width=\linewidth]{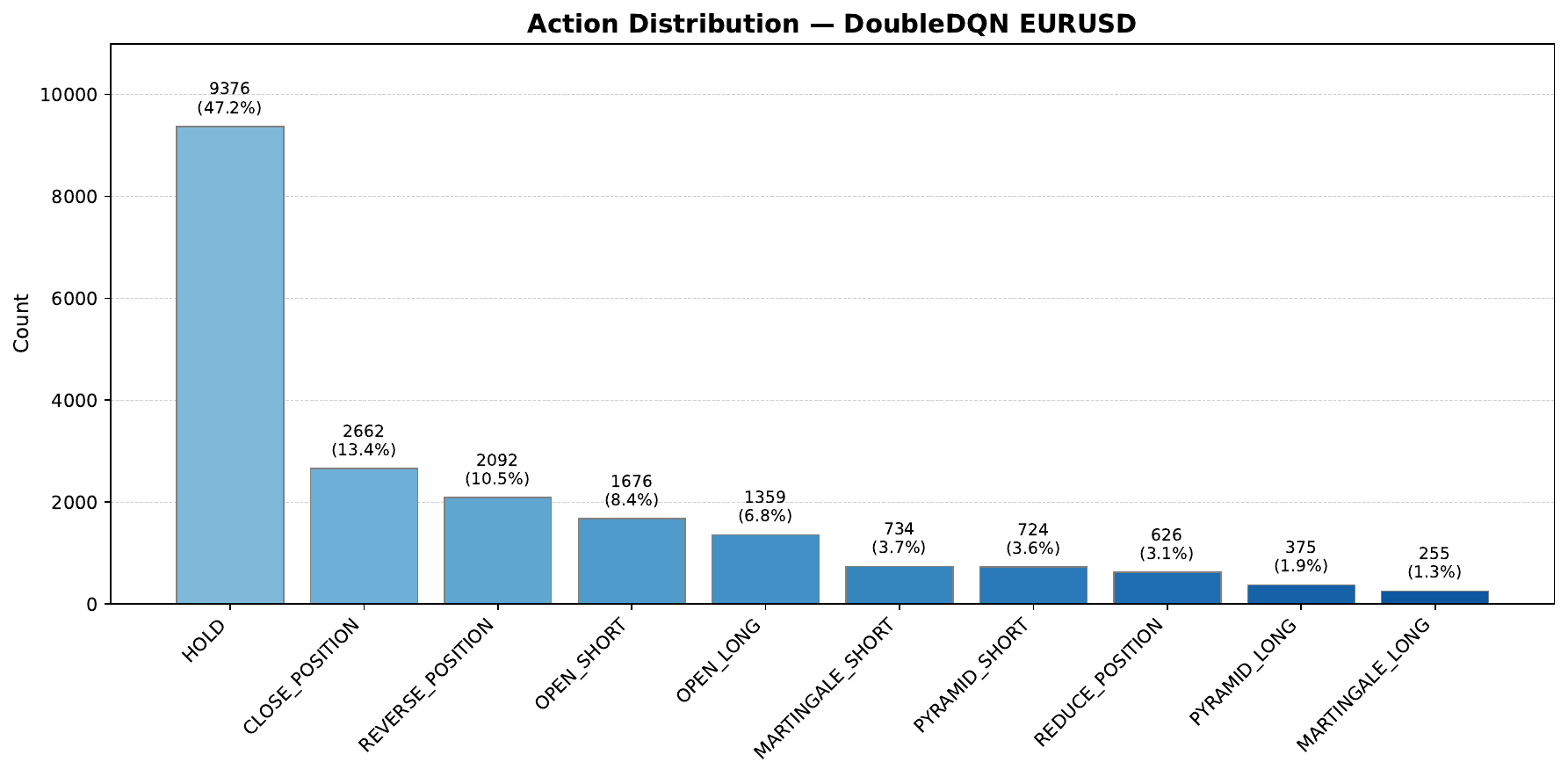}
\caption{Aggregate action distributions across training runs, showing broader action usage and higher activity in the extended interface.}
\label{fig:action_dist_agg}
\end{figure}

Endpoint metrics in \Cref{tab:action_space} suggest a return--activity trade-off, but the mechanism is clearer when inspecting action usage directly. \Cref{fig:action_dist_agg} reports aggregate action frequencies for the compared interfaces.

The distribution shows that richer action semantics are not merely available but actively exploited: scaling, reduction, and reversal primitives contribute substantial policy mass. This explains why the extended interface captures more return opportunities while simultaneously increasing turnover and path volatility.

\subsection{Scaling Mechanics Under Asymmetric Penalties}
\begin{figure}[!tbp]
\centering
\includegraphics[width=\linewidth]{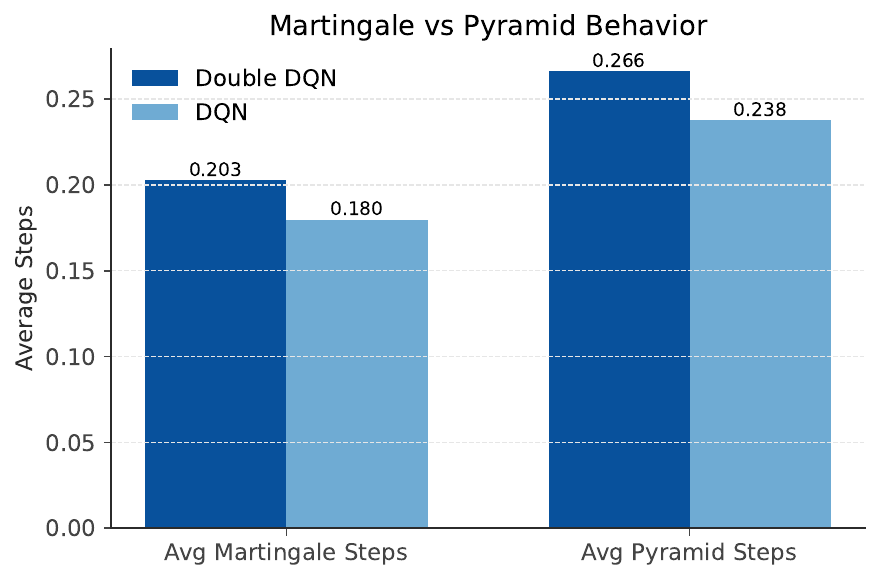}
\caption{Episode-level average pyramid depth (AvgPyr) and martingale depth (AvgMart) by scaling variant.}
\label{fig:scaling_behavior}
\end{figure}

The scaling family (\Exp{03}) is best interpreted through both endpoint values (\Cref{tab:scaling}) and operational depth statistics. \Cref{fig:scaling_behavior} visualizes average pyramid and martingale utilization by variant.

Two implications follow. First, the penalty asymmetry does shape behavior in the intended direction: enabling both mechanisms does not collapse into pure martingale usage. Second, utilization remains bounded, indicating that legality masks and margin controls are actively constraining excessive depth accumulation. The result is consistent with the design intent of making pyramiding relatively easier to sustain than martingale-style averaging down, rather than claiming that either mechanism is universally preferable.

\subsection{Interpreting Endpoint Metrics}
The retained results are intentionally restricted to the experimental scope defined in Section~\ref{sec:experiments}. Under this scope, endpoint metrics are informative about learning behavior but do not establish statistical dominance or deployment readiness. This distinction is central: the manuscript provides evidence about system dynamics, not definitive estimates of out-of-sample expected return. Consequently, directional claims should be interpreted as controlled observations that motivate multi-seed follow-up, in line with reproducibility concerns noted in prior surveys \citep{zhang2020deep}.

\subsection{Layered Safety and Causality Guarantees}
Two cross-cutting properties remain robust across experiments. First, strict anti-lookahead timing is enforced by environment mechanics rather than documentation convention, reducing leakage risk at the API level~\citep{zhang2020deep}. Second, legality masks are applied consistently during both interaction and target computation, following prior invalid-action masking practice~\citep{huang2020closer}. Together with margin and liquidation controls, this yields a defense-in-depth design appropriate for high-risk financial RL experimentation.

\subsection{Practical Interpretation}
For practitioners, the key message is to align architecture choices with deployment objectives: decomposable rewards for auditability, action-space size for desired aggressiveness, and scaling configuration for explicit return--risk preferences, consistent with classical risk-return framing in portfolio analysis~\citep{markowitz1952portfolio}. The present study provides an internally consistent, reproducible baseline for that design process; broader claims require multi-seed and out-of-sample validation.

\section{Future Work}
\label{sec:future_work}

The present manuscript establishes a controlled, reproducible baseline for Forex reinforcement learning, but several limitations naturally define the next research stage. This section outlines those limitations as concrete future-work priorities.

\subsection{Limitations of the Current Study}
All primary findings are training-split based (see Section~\ref{sec:experiments} for canonical seeding and setup). As a result, the reported differences should be interpreted as directional evidence about system behavior rather than fully characterized performance distributions. Future studies should report multi-seed confidence intervals and wider temporal stress windows to quantify stability under regime variation.

The current experiments are pair-local: each environment instance models one currency pair at a time. This excludes cross-pair capital allocation, shared margin coupling, and correlation-aware portfolio control. The cross-pair diagnostics in Section~\ref{sec:results} should therefore be read as descriptive observations, not as transfer or generalization evidence. Extending the framework to joint portfolio settings is required for full institutional realism.

\subsection{Scalability and Algorithmic Extensions}
Although the infrastructure is modular, scaling to larger instrument universes and longer histories will require more efficient replay management, distributed rollouts, and memory-aware training pipelines. From an algorithmic perspective, the release focuses on value-based baselines (DQN and DDQN) to isolate environment and reward effects. Future work should evaluate whether proximal policy optimization (PPO), soft actor-critic (SAC), and Rainbow-style extensions alter the observed reward and action-space trade-offs under identical simulator assumptions \citep{schulman2017proximal,haarnoja2018soft,hessel2018rainbow}.

\subsection{Potential Methodological Improvements}
Position management is currently discretized with fixed lot and rule-bounded scaling depth. A natural extension is hybrid control that preserves legality constraints while permitting continuous sizing and explicit risk-budget targets. Reward calibration can also be expanded by testing adaptive weighting schedules and regime-conditioned penalties, while retaining the same auditable per-component logging introduced in this paper.

\subsection{Real-World Deployment Challenges}
The current evaluation is simulation-only. It does not yet incorporate broker-specific routing behavior, execution latency, partial fills, quote outages, or production risk controls (for example, kill-switch policies and operational monitoring). Addressing these constraints requires a staged transition from historical simulation to paper trading and, eventually, limited-capital live pilots under strict governance.

In summary, future progress should combine broader uncertainty quantification, portfolio-level scaling, richer algorithmic baselines, and deployment-grade execution modeling before making strong claims about operational readiness.
\section{Conclusion}
\label{sec:conclusion}

This paper presents a modular and reproducible system-level framework for reinforcement learning-based Forex trading, centered on three design priorities: economic realism, temporal causality, and auditability. The framework combines anti-lookahead execution semantics, friction-aware portfolio accounting, decomposable reward modeling, and legal-action masking within a unified DQN-family training pipeline. It is a methodological contribution built around standard value-based learners rather than a new RL algorithm.

\paragraph{Summary of principal findings.}
Controlled experiments on EURUSD training artifacts yield three main outcomes (see Section~\ref{sec:experiments} for canonical seeding and setup):
\begin{enumerate}[leftmargin=1.8em]
    \item \textbf{Reward interactions are non-monotone.} Progressive reward composition (\variant{r1}--\variant{r7}) does not produce linear improvement. The full specification (\variant{r7}) attains the strongest endpoint profile (Sharpe $0.765$, cumulative return $57.09\%$), while intermediate variants exhibit both gains and regressions.
    \item \textbf{Action granularity induces a measurable trade-off.} The extended 10-action interface increases cumulative return and turnover, whereas the simplified 3-action adapter yields stronger conservative risk summaries (higher Sharpe and lower drawdown) in the same endpoint view.
    \item \textbf{Scaling materially improves over no scaling.} All scaling-enabled settings outperform the no-scaling baseline on drawdown. The combined configuration (\variant{s4}) achieves the best endpoint return in this artifact view, while martingale-only (\variant{s3}) attains the lowest drawdown among scaling-enabled variants.
\end{enumerate}

\paragraph{Contributions revisited.}
The core contribution is infrastructural: a research workflow in which reward decomposition, legality constraints, and anti-lookahead execution are encoded as enforceable system properties rather than informal conventions. This design improves reproducibility, supports component-level ablation, and enables more transparent diagnosis of trading-policy behavior. The contribution is therefore system-level rather than algorithmic.

\paragraph{Outlook.}
As detailed in \Cref{sec:future_work}, the next stage is to combine multi-seed uncertainty quantification, broader algorithmic baselines, and deployment-grade execution modeling. These extensions are necessary to convert controlled training evidence into robust out-of-sample and operational claims. The cross-pair diagnostics reported in Section~\ref{sec:results} remain descriptive and should not be interpreted as transfer guarantees.

\section*{Conflict of Interest}
The authors declare that they have no known competing financial interests or personal relationships that could have appeared to influence the work reported in this paper.

\printbibliography

\appendix
\section{Full Hyperparameter Configuration}
\label{app:hyperparams}

This appendix records the core configuration used across the release experiments. Main-paper values are summarized in \Cref{tab:hyperparams}; this appendix clarifies which settings were held constant and which were varied by experiment-family YAML overrides.

\subsection{Shared Training Settings}
\begin{itemize}[leftmargin=1.4em]
	\item Total timesteps: 1,000,000
	\item Replay buffer size: 40,000
	\item Batch size: 128
	\item Learn start: 10,000 steps
	\item Learn frequency: every 4 environment steps
	\item Discount factor: $\gamma=0.99$
	\item Optimizer: Adam ($2.5\times 10^{-4}$)
	\item Gradient clip: 10.0
	\item Epsilon schedule: $1.0\rightarrow0.01$ over 30,000 steps
	\item Target update interval: 2,000 learning steps
	\item Mixed precision: fp16
\end{itemize}

\subsection{Environment and Risk Settings}
\begin{itemize}[leftmargin=1.4em]
	\item Observation window: 24 bars
	\item Initial capital: USD 100,000
	\item Maximum leverage: 30x
	\item Maintenance margin ratio: 50\% of initial margin
	\item Liquidation threshold: 25\% of initial capital
	\item Commission: USD 3.5 per lot (round trip equivalent)
	\item Base deterministic slippage: 0.5 pips
	\item Rollover timestamp: 22:00 UTC (triple rollover Wednesday)
\end{itemize}

\subsection{Experiment-Family Overrides}
\begin{itemize}[leftmargin=1.4em]
	\item \Exp{01}: reward-component enable/disable schedule (\variant{r1}--\variant{r7})
	\item \Exp{02}: action-mode switch (simplified vs extended)
	\item \Exp{03}: scaling action availability (none, pyramid, martingale, both)
\end{itemize}

\subsection{Determinism Scope}
All reported runs in this manuscript used a fixed seed; see Section~\ref{sec:experiments} for the canonical seed value and related reproducibility details. This setting ensures deterministic reproducibility of the presented artifact snapshots but does not capture between-seed uncertainty.

\section{Resolved YAML Configuration Snapshots}
\label{app:configs}

This appendix provides representative resolved configurations for reproducibility. Full resolved files are stored in run artifacts under experiment output directories.

\subsection{Ablation Variant \variant{r1} (Profit-Only Core)}
\begingroup\scriptsize
\begin{verbatim}
experiment:
	family: 01_reward_ablation
	variant: r1_profit_only
agent:
	name: doubledqn
	model:
		hidden_dims: [512, 512, 256]
	training:
		total_timesteps: 1000000
		learn_start_steps: 10000
		learn_frequency: 4
reward:
	components:
		profit: {enabled: true, weight: 1.0}
		transaction: {enabled: false}
		drawdown: {enabled: false}
		volatility: {enabled: false}
		overtrading: {enabled: false}
		pyramid_penalty: {enabled: false}
		martingale_penalty: {enabled: false}
		holding: {enabled: false}
		margin: {enabled: false}
		liquidation: {enabled: false}
		constraint: {enabled: false}
training:
	random_seed: 42
\end{verbatim}
\endgroup

\subsection{Ablation Variant \variant{r7} (Full Reward)}
\begingroup\scriptsize
\begin{verbatim}
experiment:
	family: 01_reward_ablation
	variant: r7_full
agent:
	name: doubledqn
reward:
	components:
		profit: {enabled: true, weight: 1.0}
		holding: {enabled: true, weight: 0.03}
		volatility: {enabled: true, weight: 0.01}
		drawdown: {enabled: true, weight: 0.05}
		transaction: {enabled: true, weight: 0.10}
		overtrading: {enabled: true, weight: 0.02}
		pyramid_penalty: {enabled: true, weight: 0.05}
		martingale_penalty: {enabled: true, weight: 0.12}
		margin: {enabled: true, weight: 0.05}
		liquidation: {enabled: true, weight: 2.0}
		constraint: {enabled: true, weight: 0.10}
reward_normalization:
	mode: clip_only
	clip_min: -1.0
	clip_max: 1.0
training:
	random_seed: 42
\end{verbatim}
\endgroup

\subsection{Action-Space Variants}
\begingroup\scriptsize
\begin{verbatim}
# simplified
environment:
	actions:
		mode: simplified

# extended
environment:
	actions:
		mode: extended
\end{verbatim}
\endgroup

\section{Anti-Lookahead Test Specification}
\label{app:antilookahead}

This appendix formalizes the anti-lookahead guarantee enforced by the environment and summarizes corresponding validation tests.

\subsection{Formal Timing Rule}
\begin{definition}[Decision--Execution Separation]
At step $t$, the agent receives observations built from bars up to and including $\mathrm{close}_t$, where $\mathrm{close}_t$ denotes the close price of bar $t$. The chosen action $a_t$ is executed at $\mathrm{open}_{t+1}$, where $a_t$ denotes the action selected after observing $\mathrm{close}_t$ and $\mathrm{open}_{t+1}$ denotes the next bar's open used for execution. Reward and equity marking for that step are computed using information available at $\mathrm{close}_{t+1}$, where $\mathrm{close}_{t+1}$ denotes the next bar's close used for mark-to-market, without access to bar $t+2$ or later.
\end{definition}

\subsection{Validation Test Cases}
\paragraph{Test 1: Feature staleness check.}
Inject a sentinel value into a future bar and verify that the step-$t$ observation tensor does not expose this sentinel.

\paragraph{Test 2: Fill-price rule check.}
For deterministic synthetic bars, verify execution at $\mathrm{open}_{t+1}$ (not $\mathrm{close}_t$ and not $\mathrm{close}_{t+1}$).

\paragraph{Test 3: Reward-timing check.}
Verify that reward at step $t$ depends on costs and marks from bar $t+1$ only.

\paragraph{Test 4: Scaler leakage check.}
Verify that feature scaling parameters are fitted on the train split only and, if a held-out split is instantiated, reused unchanged during held-out transformation.

\paragraph{Test 5: Mask-timing check.}
Verify that legal-action masks are computed from current state/margin before action dispatch, and that illegal-action coercion/penalty is logged without future leakage.

\subsection{Interpretation}
Passing these tests does not guarantee real-market validity, but it materially reduces a major source of optimistic bias in trading RL backtests: unintended access to future information through timing or preprocessing pathways.

\section{Data and Code Availability}
\label{app:data_code_availability}

The implementation code and relevant configuration files used to reproduce the experiments in this manuscript are available in a public GitHub repository:
\url{https://github.com/NabeelAhmadSaidd/frl_trading_framework}.

Processed data artifacts used in this study are available from the corresponding author upon reasonable request.

\end{document}